\documentclass{LMCS}
\usepackage{graphicx}
\usepackage{latexsym}
\usepackage{amsfonts}
\usepackage{amsmath}
\usepackage{amssymb}
\usepackage[all]{xy}
\usepackage{array}
\usepackage{epic}
\usepackage{url}  
\usepackage{alltt}

\usepackage{enumerate,hyperref}  

\newif\ifLMCS\LMCStrue
\newif\iflncs\lncsfalse
\usepackage{amsthm}
\newcommand{\com}{\newcommand}

\newcommand{\deq}{0}
\newcommand{\deqsm}{0}
\newcommand{\dec}{-1}
\newcommand{\decsm}{-1}
\renewcommand{\vec}[1]{\mathbf{#1}}
\newcommand{\vectors}[1]{{\langle\hspace{-0.3ex}\langle#1\rangle\hspace{-0.3ex}\rangle}}

\newcommand{\bell}{\mbox{Bell}}
\newcommand{\trans}[3]{{#2,#3\models{#1}}}
\newcommand{\edge}{\leftrightarrow }
\newcommand{\procname}[1]{\textsc{#1}}  
\com{\ints}{{\mathbb Z}}
\com{\nats}{{\mathbb N}}
\let\cal=\mathcal

\com{\pgt}[1]{{\tt #1}}
\com\powerset{{\cal P}}
\com\cv{{\circ}}  
\com\clos[1]{{cl(#1)}}

\newcommand{\cat}{\cdot}   

\newcommand{\ZvsWFsym}{$\ints$}
\newenvironment{ZvsWF}{\bigskip\small\begin{list}{\ZvsWFsym:}{\leftmargin 1.5em}\item}{\end{list}}

\iflncs
\spnewtheorem{exmp}{Example}[section]{\bfseries}{\rmfamily}
\spnewtheorem{algo}{Algorithm}[section]{\bfseries}{\rmfamily}
\else
\ifLMCS
\theoremstyle{plain}\newtheorem{claim}[thm]{Claim}\newtheorem{assumption}[thm]{Assumption}
\else
\newtheorem{thm}{THEOREM}[section]
\newtheorem{obs}[thm]{OBSERVATION}
\newtheorem{cor}[thm]{COROLLARY}
\newtheorem{lem}[thm]{LEMMA}

\theoremstyle{definition}
\newtheorem{defi}[thm]{Definition}

\newtheorem{algo}[thm]{Algorithm}

\newtheorem{exa}{Example}[section]
\fi
\fi

\com{\bthm}{\begin{thm}}
\com{\ethm}{\end{thm}}
\com{\bdfn}{\begin{defi}}
\com{\edfn}{\end{defi}}
\com{\blem}{\begin{lem}}
\com{\elem}{\end{lem}}
\com{\bcor}{\begin{cor}}
\com{\ecor}{\end{cor}}
\ifLMCS
\com{\bex}{\begin{exa}}
\com{\eex}{\par\bigskip\end{exa}}
\else
\com{\bex}{\begin{exa}\pushQED{\qed}}
\com{\eex}{\popQED\end{exa}}
\fi
\com{\bprf}{\begin{proof}}
\iflncs
\com{\eprf}{\qed\end{proof}}
\else
\com{\eprf}{\end{proof}}
\fi

\newcommand{\bi}{\begin{enumerate}[$\bullet$]}
\newcommand{\ei}{\end{enumerate}}
\newcommand{\be}{\begin{enumerate}[(1)]}
\newcommand{\ee}{\end{enumerate}}

\com{\fl}{\noindent}
\com{\hair}{\hspace{3mm}}
\com{\vair}{\vspace{3mm}}

\newenvironment{fig0}[3]
{
\xdef\fighack{\noexpand\caption{#1}\noexpand\label{#3}}
\begin{figure*}[#2]
\hspace*{3mm}\begin{minipage}{0.95\textwidth}}{\end{minipage}%
\fighack%
\end{figure*}}

\newenvironment{program}{
\medskip\par
\begin{minipage}{0.9\textwidth}
\begin{alltt}}{
\end{alltt}
\end{minipage}\par\medskip\par}

\title{Monotonicity Constraints for Termination in the Integer Domain}

\iflncs
\institute{Academic College of Tel-Aviv Yaffo}
\else
\ifLMCS
 \author[A. M. Ben-Amram]{Amir M. Ben-Amram}
 \address{School of Computer Science, The Tel-Aviv Yaffo Academic College, Israel}	
 \email{benamram.amir@gmail.com}  

\else
\author{Amir M.\ Ben-Amram}
\fi

\def\doi{7 (3:04) 2011}
\lmcsheading%
{\doi}
{1--43}
{}
{}
{Apr.~26, 2010}
{Aug.~24, 2011}
{}

\begin{document}

\maketitle

\begin{abstract}
Size-Change Termination (SCT) is a method of proving
program termination based on the impossibility 
of infinite descent. To this end we use a program abstraction in which
transitions are described by
\emph{monotonicity constraints} over (abstract) variables.
When only constraints of the form $x>y'$ and $x\ge y'$ are allowed, we have
size-change graphs. In the last decade, both theory and practice have
evolved significantly in this restricted framework. The crucial underlying assumption
of most of the past work is that the domain of the variables is well-founded.
In a recent paper I showed how to extend and adapt some theory from
the domain of size-change graphs to general monotonicity constraints, thus complementing
previous work, but remaining in the realm of
well-founded domains.
However, monotonicity constraints are, interestingly, capable of proving termination
also in the integer domain, which is not well-founded.

The purpose of this paper is to explore the application of monotonicity constraints
in this domain.
We lay the necessary theoretical foundation, and present precise decision procedures for
termination; finally, we provide a procedure to construct explicit
global ranking functions from monotonicity constraints
in singly-exponential time, and of optimal worst-case size and dimension (ordinal).
\end{abstract}

\section{Introduction}
This paper concerns automated termination analysis---deciding whether a program
 terminates, and possibly generating a \emph{global ranking function}.
The termination problem is well-known undecidable for Turing-complete programming languages;
one of the ways in which this obstacle may be circumvented is to study, for this purpose,
a class of \emph{abstract programs} that are expressive
enough to allow many concrete programs to be represented, so that termination
of the abstraction implies termination of the concrete program. An important point is that the abstract programs do
not constitute a Turing-complete programming language and \emph{can}
have a decidable termination problem.  The abstraction studied in this paper
is \emph{monotonicity constraint transition systems}~\cite{Codish-et-al:05,BA:mcs}.
The term will be usually abbreviated to MCS.

The MCS abstraction is an extension of the SCT
(Size-Change Termination~\cite{leejonesbenamram01})  abstraction, which has been studied quite
extensively during the last decade%
\footnote{References are too numerous to cite here, but see
the author's summary web page \url{http://www2.mta.ac.il/~amirben/sct.html}).}%
.
represent a program as a transition system with states. Abstraction of a program
consists of the formation of a \emph{control-flow graph}
for the program, identification of a set of \emph{state variables},  and formation of a finite set of 
abstract transitions (i.e., abstract descriptions of program steps, where the notion of step can be tuned to different
needs).

In the SCT abstraction, an abstract transition is specified by a set of inequalities, that relate variable
values in the target state to those
of the source state (these inequalities are often represented by a \emph{size-change graph}).
Extending this notion, 
a \emph{monotonicity constraint} (MC) allows for any
conjunction of order relations, including
equalities as well as strict and non-strict inequalities, and involving 
any pair of variables from the source state and target state. 
The \emph{Monotonicity Constraint Transition Systems} treated in this paper also allow
constraints to be associated with a point in the
control-flow graph (these are called \emph{state invariants}).

The size-change technique was conceived to deal with
well-founded domains, where infinite descent is impossible.
Termination is deduced by proving that any (hypothetical) infinite run would
decrease some value monotonically and endlessly, so that well-foundedness would be
contradicted.

Monotonicity constraints generalize the SCT abstraction
and are clearly more expressive, a fact that was highlighted by
Codish, Lagoon and Stuckey~\cite{Codish-et-al:05}. 
They made the 
intriguing observation that pre-existing termination analyzers based on monotonicity constraints
\cite{LS:97,CT:99,LSS:04} apply a termination test which is
sound and complete for SCT, but incomplete for general monotonicity constraints,
even if one does not change the underlying model---namely that ``data'' are from an
unspecified well-founded domain. In addition, they pointed out that 
monotonicity constraints can imply termination under a different assumption---that
the data are integers. 
 Integers, not being well-founded, cannot be handled by the SCT abstraction. 

In practice, the integers are already the predominant domain for
monotonicity constraints and size-change termination.
Often---in particular in functional and logic programming---they represent the size of a list or
 tree (whence the term \emph{size-change termination}), and
 are necessarily non-negative, which allows the well-founded model to be 
used. In contrast, in certain application domains, and typically in imperative programming,
the crucial variables are of integer type and might be negative
(by design or by mistake).
 But MCs can still imply termination, as witnessed by the loop
\verb/while(x<y) x=x+1/.  The value of \pgt{x} does not descend but grow; however 
the constraint $\pgt{x}<\pgt{y}$ (along with the fact that \pgt{y} does not change)
tells us that this cannot go on forever.

In a previous paper~\cite{BA:mcs}, the theory of monotonicity constraint transition systems
in the well-founded model was investigated. Main results include:

\begin{enumerate}[$\bullet$]
\item
 The syntax and semantics of the abstraction are presented.
\item
 A \emph{combinatorial} termination criterion is formulated in terms of the
representation of monotonicity constraints as graphs
(briefly: the existence of an infinite descending path, or walk, in every infinite multipath).
This is an adaptation of the SCT criterion from~\cite{leejonesbenamram01}.
\item
 It is proved that satisfaction of this criterion is equivalent to the termination of transition system that satisfies
 the constraints---in logical terms, the criterion is sound and complete.
\item
Termination of MCSs is shown decidable, and more precisely,
PSPACE-complete.
Two decision procedures are given: a direct \emph{closure-based} algorithm,
and a reduction of the problem to SCT, which provides an alternative algorithm, along with insight into the relationship
between the two constraint domains.
\item
An algorithm is given to construct explicit
global ranking functions given any terminating MCS
(recall that a global ranking function is a function from program states into a well-founded domain,
that decreases on every transition). The algorithm has optimal time complexity ($2^{O(n\log n)}$) and produces
a ranking function whose values are tuples, under the lexicographic ordering. It is also optimal in 
the dimension (length of the tuples, implying the ordinal of the codomain), at least in a worst-case sense.\medskip
\end{enumerate}

\noindent The contribution of this paper is to obtain some similar results in the integer model.
This development is more complicated than in the well-founded model,
but is certainly worthwhile due to the
practical importance of this domain (many published termination analyses
target the integers specifically; a few, closest to this work, are
referenced in the related-work section).

An intuitive reduction of the integer case back to the well-founded case is to create
a new variable for every difference $x_i-x_j$ which can be deduced from the
constraints to be non-negative, and also deduce relations among these new variables to obtain
a new abstract program over the natural numbers.
But this solution may square the
number of variables, which is bad because this number is in the exponent of the
complexity, and in general is not complete (see Example~\ref{ex:int3}) in the next section).
 We tackle the problem directly instead, but as an interesting corollary we shall find
that the above reduction is, in fact, a correct solution given a certain preprocessing of the program.

This paper is organized as follows.  In the next section we formally introduce
monotonicity constraint transition systems and their semantics.  After this quick
technical introduction a sample of examples are given, witnessing the range of 
termination arguments captured by this framework.
Section~\ref{sec:stable} recalls the notion of a \emph{stable} MCS (introduced in~\cite{BA:mcs}).
Stabilization propagates invariants around the abstract program and makes it more amenable to local analysis.
Section~\ref{sec:walks} gives a combinatorial termination criterion (in terms of graphs), similar to what has been
known for SCT and for MCS in previous work. The criterion is proved sound and complete and decision algorithms
are discussed. In Section~\ref{sec:elaborate}, we recall another notion from~\cite{BA:mcs}, that of \emph{elaboration}.
Elaborating a systems makes some information which implicit in it, explicit, and further simplifies its algorithmic
processing. It is used in Section~\ref{sec:grf}, presenting the algorithm to construct ranking functions.
Section~\ref{sec:rooted} complements the previous sections by briefly explaining the role of reachability (sometimes
a program may seem to contain an infinite loop, but it is not reachable). Section~\ref{sec:rw} discusses related
work and Section~\ref{sec:conclusion} concludes.

This paper is intended to be self-contained, so that it can be read independently of its predecessor~\cite{BA:mcs}.
However, a reader may be interested to know what the challenges were in handling the integer domain in constrast with 
the well-founded domain, where the definitions or techniques are the same, when they differ and how. 
The paper is interspersed with special comments marked with the symbol {\ZvsWFsym}.
These comments are meant to answer the above questions. They can be skipped without compromising the integrity of the text.

The central results of this work have been presented in the 20th International Conference on Computer Aided Verification conference (2009)
 and are stated, very briefly, in the proceedings~\cite{BA:cav09} .

\section{Basic Definitions and Motivating Examples}
\label{sec:ints}

\begin{ZvsWF}
The basic definitions (Sect.~\ref{sec:MCSdef}--\ref{sec:MCSsem}) are essentially as in~\cite{BA:mcs},  except for the notation
$\pi$-termination, introduced for distinction between the integer interpretation of an MCS and the well-founded interpretation.
Section~\ref{sec:examples} includes examples to show
the expressiveness of the integer MCS model and contrast it with the usage of SCT to analyze the same programs.\end{ZvsWF}

\subsection{Monotonicity constraint transition systems}
\label{sec:MCSdef}

A monotonicity constraint transition system is an abstract program.
An abstract program is, essentially, a set of \emph{abstract transitions}. An
abstract transition is a relation on (abstract) program states. 
When describing program transitions, it is customary to mark the variables
in the \emph{resulting} state with primes (e.g., $x'$). 
For simplicity, we will name the variables $x_1,\dots,x_n$ (regardless of
what program variables they represent). This notation also suggests that the same
number of variables ($n$) is used to represent all states. Of course, in actual programs
this is not necessarily the case (for example due to different scopes) and it will be
more efficient to maintain only the necessary variables at each program point.

\bdfn[MCS]
A \emph{monotonicity constraint transition system}, or MCS, is an abstract
program representation that consists of a control-flow graph (CFG),
monotonicity constraints and state invariants, all defined below.
\begin{enumerate}[$\bullet$]
\item
A control-flow graph is a directed graph (allowing parallel arcs)
over the set $F$ of \emph{flow points}.

\item
A \emph{monotonicity constraint} (MC) 
is a conjunction of 
order constraints $x \bowtie y$ where
 $x,y\in\{x_1,\dots,x_n,x_1',\dots,x_n'\}$, and ${\bowtie}\in\{>,\ge,=\}$.

\item
Every CFG arc $f\to g$  is associated with a
monotonicity constraint $G$. We write $G:f\to g$.

\item
For each $f\in F$, there is an \emph{invariant} $I_f$, which is
a conjunction of order constraints among the variables.
\end{enumerate}
\edfn

\noindent The terms ``abstract program"  and ``MCS" are used
interchangeably, when context permits. The letter $\cal A$ is usually used
to denote such a program; $F^{\mathcal A}$ will be its flow-point set. When notions of 
connectivity are applied to $\cal A$ (such as, ``$\cal A$ is strongly connected''),
they concern the underlying CFG.

\subsection{Semantics}
\label{sec:MCSsem}

\bdfn[states]
A {\em state\/} of $\cal A$ (or an abstract state)
is $s=(f,\sigma)$, where
$f\in F^{\mathcal A}$ and $\sigma:\{1,\dots,n\}\to\ints$ represents
 an assignment of values
to the variables.
\edfn

Satisfaction of a predicate $e$ with free variables $x_1,\dots,x_n$ (for example, $x_1>x_2$) by
an assignment $\sigma$ is defined in the natural way, and expressed by
$\sigma\models e$.
If $e$ is a predicate involving the $2n$ variables $x_1,\dots,x_n, 
x_1',\dots,x_n'$, we write $\sigma,\sigma' \models e$ when $e$ is
satisfied by setting the unprimed variables according to
$\sigma$ and the primed ones according to $\sigma'$.

\bdfn[transitions] \label{def:trans}
 A transition is a pair of states, a \emph{source state} $s$ and a \emph{target state} $s'$.
For $G:f\to g\in {\cal A}$, we write $\trans{G}{(f,\sigma)}{(g,\sigma')}$ if
$\sigma\models I_f$, $\sigma'\models I_g$ and $\sigma,\sigma' \models G$.
We say that transition $s\mapsto s'$ satisfies $G$.
 \edfn
 

Note that we may have unsatisfiable MCs, such as $x_1>x_2
\land x_2>x_1$; it is useful to view all such MCs as synonyms and use
the common notation $\bot$ for them (as one would typically do in an
Abstract Interpretation domain).

\bdfn[transition system]
The \emph{transition system} associated with ${\cal A}$
is the binary relation $$T_{\cal A} = \{ (s,s') \mid \trans{G}{s}{s'} 
\text{ for some }G\in {\cal A}\}.$$
\edfn

\noindent Note that a program representation may also be called a
``transition system.'' I am using the unqualified term for a semantic
object. The program representation is referred to as an MC transition
system (or sometimes just MC system).

\bdfn[run]
A {\em run\/} of ${\cal T}_{\cal A}$ is a (finite or infinite) sequence
of states $\tilde s = s_0,s_1,s_2\dots$ such that for all $i$,
 $(s_i, s_{i+1})\in {\cal T}_{\cal A}$.
\edfn

Note that by the definition of ${\cal T}_{\cal A}$, a run is associated
with a sequence of CFG arcs labeled by $G_1,G_2,\dots$ where
$\trans{G_i}{s_{i-1}}{s_i}$. This sequence constitutes a (possibly non-simple) path in the CFG.

\bdfn[termination]
Transition system ${\cal T}_{\cal A}$ is {\em uniformly terminating\/} if
it has no infinite run.  MCS $\cal A$ is said to be \emph{$\pi$-terminating} if
${\cal T}_{\cal A}$ is uniformly terminating.  
\edfn

The prefix $\pi$  (a symbol for the order type of the integers~\cite{Schaum:sets})
is included since the same MCS can be interpreted in the well-founded model of~\cite{BA:mcs} and may possibly
be non-terminating there (though this would require the value domain to be of an order type greater
than $\omega$). The term \emph{uniform} refers to termination that is independent of the initial state.
In Section~\ref{sec:rooted} we discuss the consequences of specifying an initial flow-point.

\paragraph*{Remark.}
It may be tempting to abstract away from the integers and, just as the well-founded
case was treated generally (so that any well-founded domain could be used),
give a general definition of the property that the domain has to satisfy
for our termination arguments to apply. The property is that for any two elements there are
 only finitely many elements strictly between them.
However, this abstraction buys us no generality, as every total order with this property
is isomorphic to a subset of the integers. On the other hand, the fact that
we assume the underlying set to be, specifically, the integers, has the
satisfactory outcome that \emph{completeness} of our deduction means that 
specific knowledge about the integers could not have been used in any better way.

\subsection{Examples}
\label{sec:examples}

Here are a few small examples that illustrate the transformation of programs into MCS.
The reader is invited to verify that the abstract programs created are indeed terminating.
To contrast the expressiveness of the MC abstraction with that of the better-known SCT,
comments are included regarding how the examples might be treated by SCT-based tools.
Recall that in the SCT framework, only well-founded domains are handled.

\bex\label{ex:int1}
Consider the following program, in a self-explanatory functional programming
language with integer data.

\begin{program}
f(m,n) = if m <= 0 then n
    else if n <= 0 then f(m-1, n-1)
                   else f(m, n-1)
\end{program}

\noindent In this example the recursive calls depend on some variable being positive,
and therefore the proof can somehow be embedded in the well-founded framework.
This requires placing the control-flow points at the call sites, where we can
rely on the guards; it is also necessary to deal with the fact that \pgt{n},
while being pertinent to the termination proof, is only known to be positive
(and thus can be admitted into the SCT abstraction) in the second call.
When using MCs, we need not worry about all of this.  We can place a single
flow-point at the function entry (as done in the simple-minded abstraction
of~\cite{leejonesbenamram01}).  When analysing the body of the function, we
create an abstract transition for each recursive branch, taking into account the
branch condition. The abstract variables will
be {\tt m}, {\tt n} and {\tt 0} (the analyser should recognize that these
three quantities are involved in the comparisons).
The resulting MCS thus consists of two abstract transitions:
\begin{align*}
G_1&:\quad
 \pgt{m}>\pgt{0}\land \pgt{n}>\pgt{0}\land \pgt{n}>\pgt{n}'\land \pgt{m}=\pgt{m}'\land \pgt{0}=\pgt{0}' \\
G_2&:\quad
 \pgt{m}>\pgt{0}\land \pgt{n}\le\pgt{0}\land \pgt{n}>\pgt{n}'\land \pgt{m}>\pgt{m}'\land \pgt{0}=\pgt{0}'   \qedhere\\
\end{align*}
\eex

\bex\label{ex:int2}
This simple example involves descent in a difference.

\begin{program}
s(m,n) = if m > n then 1 + s(m-1, n)
                  else 0
\end{program}

\noindent Here, it is possible to use the SCT framework by placing the control-flow
point at the call site, where we know that the difference $\pgt{m}-\pgt{n}$ is positive,
and using the difference as the abstract variable. With an MCS, we
can use a straight-forward translation:
\begin{align*}
G_1&:\quad
 \pgt{m}>\pgt{n}\land \pgt{n}=\pgt{n}'\land \pgt{m}>\pgt{m}' 
\qedhere\\
\end{align*}
\eex

\noindent As noted, both of the above examples could be proven
terminating by SCT using an abstraction that relies on invariants and
judicious placement of the flow-points; such techniques have been
implemented, for example, in the theorem prover ACL2~\cite{MV-cav06}.
The next programs challenge to the simple invariant-based technique,
and therefore illustrate more significantly the advantage of
monotonicity constraints.

\bex\label{ex:int3}
Consider the following program:

\begin{program}
g(m,n) = if m <= 0 then n
                   else g(n, m-1)
\end{program}

\noindent Both parameters are pertinent to termination, but only
\pgt{m} is known to be positive at the call site.  But as $\pgt{m}'$
is unrelated to $\pgt{m}$, this does not suffice to prove
termination. Neither is the difference $\pgt{n}-\pgt{m}$ useful as an
abstract variable, since there is no information on its sign and it
does not change monotonically. Nonetheless, it is not hard to prove
termination with the straight-forward MCS abstraction:
\begin{align*}
G_1&:\quad
 \pgt{m}>\pgt{0}\land \pgt{n}=\pgt{m}'\land \pgt{m}>\pgt{n}' 
\qedhere
\end{align*}
\eex

\bex\label{ex:int_and_bool}
In this example,
unsatisfiability of certain paths is crucial to the termination proof.

\begin{program}
while (0<x<n)
   if b then
    x := x+1
   else
    x := x-1
\end{program}

\noindent A way to handle this program with SCT might be
the calling-context method implemented in
ACL2; the theorem prover is used to discover the fact that an increment cannot be followed
by a decrement (or vice versa). It produces an abstract program into which this information is
already coded. It also produces
the abstract variable $\pgt{n}-\pgt{x}$.
In contrast,
with monotonicity constraints, a na\"ive translation of the program is sufficient.
Since only the integer domain is treated in this work, we represent the Boolean variable
\pgt{b} as an integer with the test interpreted as \pgt{b>0}.
This yields:
\begin{align*}
G_1&:\quad
 \pgt{0}<\pgt{x}\land \pgt{x}<\pgt{n}\land \pgt{b}>\pgt{0} 
 \land \pgt{x}<\pgt{x}' \land \pgt{n}=\pgt{n}' \land \pgt{b}=\pgt{b}'
 \land \pgt{0}=\pgt{0}' \\
G_2&:\quad
 \pgt{0}<\pgt{x}\land \pgt{x}<\pgt{n}\land \pgt{b}\le \pgt{0} 
 \land \pgt{x}>\pgt{x}' \land \pgt{n}=\pgt{n}' \land \pgt{b}=\pgt{b}'
 \land \pgt{0}=\pgt{0}' 
\qedhere
\end{align*}
\eex

\bex \label{ex:addition}
Consider the following program.

\begin{program}
g(m,n) = if m <= 0 then n
                   else g(m+n, n-1)
\end{program}

\noindent The apparent difficulty here is that the sign of \pgt{n} is
not known at the point where it is added to \pgt{m}. Hence, one cannot
adequately represent the effect of such an update by a monotonicity
constraint. A similar problem arises with other operations (notably
subtraction).  However, this nut is easily cracked, since one
\emph{can} represent the effect of addition as a disjunction of three
monotonicity constraints. Thus, for the above program, we have the
following (terminating) abstraction:
\begin{align*}
G_1&:\quad
 \pgt{m}>\pgt{0}\land \pgt{n}>\pgt{0}\land\pgt{m}<\pgt{m}'\land \pgt{n}>\pgt{n}'\land \pgt{0}=\pgt{0}'\\
G_2&:\quad
 \pgt{m}>\pgt{0}\land \pgt{n}=\pgt{0}\land\pgt{m}=\pgt{m}'\land \pgt{n}>\pgt{n}'\land \pgt{0}=\pgt{0}'\\
G_3&:\quad
 \pgt{m}>\pgt{0}\land \pgt{n}<\pgt{0}\land\pgt{m}>\pgt{m}'\land \pgt{n}>\pgt{n}'\land \pgt{0}=\pgt{0}'
\qedhere
\end{align*}
\eex

\noindent Note that when $n$ is initially positive, the first parameter grows at first, until $n$ reaches zero and then
the first parameter begins to shrink. Programs with such ``phase shift'' have attracted the attention
of termination researchers, and appear in several publications, e.g.,~\cite{BMS:Polyranking}.

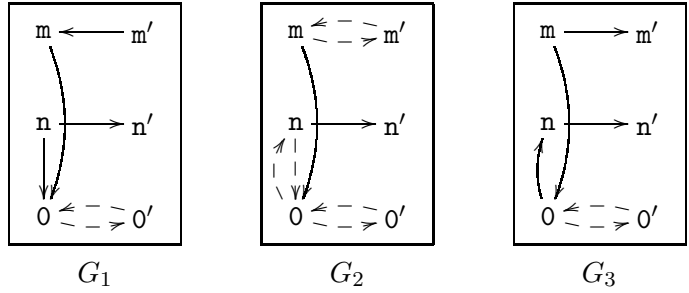
\begin{fig0}{MCs of Example~2.11 as graphs.
The left-hand side is the source. Broken arcs represent non-strict descent.}{t}
{fig:exintro}
\begin{centering}
\setlength{\extrarowheight}{1ex}
\begin{tabular}{@{\extracolsep{20pt}}ccc}
\fbox{\ $\xymatrix@R=20pt{
  \texttt{m}\ar@/^8pt/@{->}[dd]& \texttt{m}'\ar@{->}[l]  \\
  \texttt{n}\ar@{->}[d]\ar@{->}[r] & \texttt{n}' \\
  \texttt{0}\ar@/_4pt/@{-->}[r]    & \texttt{0}'\ar@/_4pt/@{-->}[l]
}$\ } &
\fbox{\ $\xymatrix@R=20pt{
  \texttt{m}\ar@/_4pt/@{-->}[r]\ar@/^8pt/@{->}[dd]& \texttt{m}'\ar@/_4pt/@{-->}[l]  \\
  \texttt{n}\ar@{-->}[d]\ar@{->}[r] & \texttt{n}' \\
  \texttt{0}\ar@/_4pt/@{-->}[r]\ar@/^8pt/@{-->}[u] & \texttt{0}'\ar@/_4pt/@{-->}[l]
}$\ } &
\fbox{\ $\xymatrix@R=20pt{
  \texttt{m}\ar@{->}[r]\ar@/^8pt/@{->}[dd]& \texttt{m}' \\
  \texttt{n}\ar@{->}[r] & \texttt{n}' \\
  \texttt{0}\ar@/_4pt/@{-->}[r]\ar@/^4pt/@{->}[u] & \texttt{0}'\ar@/_4pt/@{-->}[l]
}$\ } \\
$G_1$ & $G_2$ & $G_3$ \\
\end{tabular}\par
\end{centering}
\end{fig0}

\subsection{A comment regarding state invariants}

None of the above examples used state invariants (associated with a flow-point
rather than a transition), and in fact it is easy to see that one can always do
without them,
as it is
possible to include the constraints $I_f$ in every MC that transitions from $f$.
However, it may be convenient to make the
association of certain assertions with a flow point, rather than a 
specific transition, explicit, and our algorithms make significant use of such
invariants.

\begin{ZvsWF}
The next two sections deal, respectively, with the representation and transformation of MCSs;
the ideas are not essentially different from those used in~\cite{BA:mcs} but there are some new details
which are important for the sequel.
\end{ZvsWF}

\subsection{Weighted graph representation}

An MC can be represented by a labeled digraph (directed graph). 
This representation enables a style of reasoning, using graph properties like
paths, which has been very useful in most, if not all, previous work on SCT and
monotonicity constraints.
An arc $x\xrightarrow{r} y$ ($r$ is the label) represents a relation 
$x>y$ or $x\ge y$. One only needs two labels, $>$ and $\ge$.
It is convenient to use integers for the labels, and apply techniques
from the world of weighted graphs. Hence, we employ the term \emph{weighted graph representation}.

\bdfn
The weighted graph representation of a monotonicity constraint is a weighted digraph 
with node set 
$\{x_1,\dots,x_n,x_1',\dots,x_n'\}$ and 
  for each constraint $x>y$ (respectively $x\ge y$), an arc $x\xrightarrow{\decsm} y$
($x\xrightarrow{\deqsm} y'$).
The arcs are referred to, verbally, as \emph{strict} (label $\dec$) or
\emph{non-strict} (label $\deq$).
\edfn

The notation $x\to y$ may be used to represent an arc from $x$
to $y$ (of unspecified label). In diagrams, to avoid clutter, we distinguish
the types of arcs by using a dashed arrow for the weak inequalities
(see Figure~\ref{fig:exintro}).
Note that an equality constraint $x=y$ is represented by a pair of non-strict arcs. In certain algorithms,
it is convenient to assume that the such arcs are distinguished from ``ordinary''
non-strict arcs. We refer to them as \emph{no-change arcs}.

\bdfn[$\vdash$]  \label{def:vdash}
Let $P(s,s')$ be any predicate over
states $s,s'$, possibly written
using variable names, e.g., $x_1>x_2 \land x_2<x_2'$.
We write $G\vdash P$ if
$\forall s,s': \trans{G}{s}{s'} \Rightarrow P(s,s')$.
\edfn

\bdfn
A monotonicity constraint $G$, in graph representation, is \emph{closed under logical consequence}
(or just closed) if, whenever
$G\vdash x > y,$ or $G\vdash x \ge y,$ for 
$x,y\in \{x_1,\dots,x_n, x_1',\dots,x_n'\}$,
the stronger of the implied relations is explicitly included in the graph.
\edfn

Note that for $G:f\to g$,
the condition $G\vdash P$ takes the invariants $I_f$
and $I_g$ into account (consider Definitions~\ref{def:trans} and~\ref{def:vdash}).
Thus, a closed MC subsumes the invariants in its source and target
states.

\bdfn
An MC $H$ \emph{is at least as strong as} $G$
if whenever $G\vdash P$, also $H\vdash P$.
The (consequence) \emph{closure} of a monotonicity constraint $G$,
denoted $\overline G$, is 
the weakest MC
that is at least as strong as $G$ and is consequence-closed.
\edfn

The closure of a graph be computed efficiently ($O(n^3)$ time) by
a DFS-based algorithm. First the graph is divided in strongly connected components.
Within every component, only non-strict arcs should appear; otherwise the graph is
unsatisfiable and should be immediately replaced by the fixed object $\bot$.
Otherwise, the components represent groups of variables constrained to be equal, and
the acyclic graph of components can be processed to determine the constraints relating each pair.

Henceforth, we identify a MC with its consequence-closed graph representation,
and assume that this is how our algorithms will maintain them. 
In graph representation, the restriction of the SCT framework is that only arcs of the
form $x\to y'$ are admitted. Graphs of this form are called \emph{size-change graphs} in the
SCT literature and so will they be called in this paper.

\subsection{Transforming MC systems}
\label{sec:transforming}

A key tool in processing MC systems is the idea of transforming them for the purposes of analysis.
We employ two kinds of transformations: the first kindcreates a new system that 
is ``equivalent'' in the sense that they have the same runs, up to renaming of flow-points and possibly variables---the precise
notion is \emph{bisimulation}, defined below. The second kind of transformation 
creates a system that represents \emph{part} of the runs of the
given one---see the definition of \emph{restriction} below.  These definitions are given here for later reference. The first
(bisimulation) is cited verbatim from~\cite{BA:mcs}.

\bdfn
Let $\cal A$, $\cal B$ be transition systems, with flow-point sets $F^{\cal A}$,
$F^{\cal B}$ respectively, 
and both having states described by $n$ variables.
We say that $\cal A$ 
\emph{simulates}
$\cal B$  if there is a relation 
$\phi\subseteq F^{\cal B}\times F^{\cal A}$ (``correspondence of flow-points")
and, for all $(f,g)\in\phi$, a bijection
$\psi_{g,f}:  \{1,\dots,n\}\to\{1,\dots,n\}$  (``variable renaming")
 such that for every (finite or infinite) state-transition sequence
 $(f_1,\sigma_1)\mapsto (f_2,\sigma_2)\mapsto (f_3,\sigma_3)\mapsto \dots$ 
of $\cal B$ there is a corresponding sequence
$(g_1,\sigma'_1)\mapsto (g_2,\sigma'_2)\mapsto (g_3,\sigma'_3)\mapsto \dots$
of ${\cal A}$ with $(f_i,g_i)\in \phi$ and
$\sigma'_i = \sigma_i\circ(\psi_{g_i,f_i})$.
We say that $\cal A$ 
\emph{bisimulates} $\cal B$ if, in addition,
 for every (finite or infinite) state-transition sequence
$(g_1,\sigma'_1)\mapsto (g_2,\sigma'_2)\mapsto (g_3,\sigma'_3)\mapsto \dots$
of $\cal A$ there is a corresponding sequence
$(f_1,\sigma_1)\mapsto (f_2,\sigma_2)\mapsto (f_3,\sigma_3)\mapsto \dots$
of ${\cal B}$, also with $(f_i,g_i)\in \phi$ and
$\sigma'_i = \sigma_i\circ(\psi_{g_i,f_i})$.
\edfn

Thus, $\cal A$ bisimulates $\cal B$ if they simulate each other via the same
pair of mappings.

\bdfn
We say that an abstract program $\cal A$ (bi-)simulates an abstract program $\cal B$
if ${\cal T}_{\cal A}$ 
(bi-)simulates
${\cal T}_{\cal B}$, via mappings $\phi$ and $\psi$, as above.

We say that $\cal A$ simulates $\cal B$ \emph{deterministically} if for every
$f\in F^{\cal B}$ and assignment $\sigma$ satisfying $I_{f}$
there is a unique $g\in F^{\cal A}$ with
$(f,g)\in \phi$ such
that, letting $\sigma' = \sigma\circ(\psi_{g,f})$, assignment $\sigma'$ satisfies $I_{g}$.

If  $\cal A$ bisimulates $\cal B$, and  $\cal A$ simulates $\cal B$
deterministically, we say (for brevity) that $\cal A$ bisimulates
$\cal B$ deterministically.
\edfn

Determinism means that
the invariants of different $\cal A$
flow-points that simulate a given $\cal B$ flow-point have to
be mutually exclusive.

\bdfn
The notation $\mathcal A \sim_{\phi,\psi} \mathcal B$ means that $\cal A$ simulates $\cal B$ deterministically via
flow-point correspondence $\phi$ and the variable renaming function $\phi$. We omit $\phi$ when it is the identity,
and omit both if it is not important to specify them.
\edfn

\bdfn
Let $\cal A$, $\cal B$ be transition systems, with same flow-point sets $F^{\cal A} = F^{\cal B}$, 
and both having states described by $n$ variables.
We say that $\cal A$ 
\emph{is a restriction of} $\cal B$, and write $\mathcal A \Subset \mathcal B$, if
every (finite or infinite) state-transition sequence
 $(f_1,\sigma_1)\mapsto (f_2,\sigma_2)\mapsto (f_3,\sigma_3)\mapsto \dots$ 
of $\cal B$ is also a transition sequence of ${\cal A}$.
\edfn

A restriction results, obviously, from tightening the constraints at certain flow-points or transitions.

\bdfn
We write $\mathcal A \Subset_{\phi} \mathcal B$, if there exists $\mathcal C$ such that
$\mathcal A \Subset \mathcal C$ and $\mathcal C \sim_{\phi} \mathcal B$.
\edfn

\section{Stable Systems}
\label{sec:stable}

The notion of \emph{stable} MC systems 
is from~\cite{BA:mcs}. This section recalls the definition
and states some consequences of stability that are used in forthcoming proofs.

\begin{ZvsWF} There is no novelty here.
\end{ZvsWF}

\bdfn
An MCS $\cal A$ is \emph{stable} if 
(1) all MCs in $\cal A$ are satisfiable;
(2) For all $G:f\to g$ in $\cal A$, whenever
$G\vdash  x_i \rhd x_j$ (some relation between source variables),
also $I_f\vdash x_i \rhd x_j$.
(4) Similarly, if
$G\vdash   x_i' \rhd x_j',$
also $I_g\vdash x_i \rhd x_j$.
\edfn

Note that while consequence-closure requires that all information from $I_f$ and $I_g$
be present in $G$, stability requires that $G$ cannot add
information to $I_f$ and $I_g$.

Stabilizing a given system may require flow-points to be duplicated, since two
MCs coming out of $f$ may disagree on the conditions that must be placed
in $I_f$.

It is always possible to tranform an MC system $\mathcal B$ into a stable $\mathcal A$ such that
$\mathcal A \sim \mathcal B$; this transformation is called \emph{stabilization},
and an algorithm is described
in~\cite{BA:mcs}.
Stabilization does not rename variables, but may require flow-points to be duplicated, since two
MCs coming out of $f$ may disagree on the conditions that must be placed
in $I_f$.
 In the worst case, it can multiply the size of the system
by a factor exponential in the number of variables $n$
(specifically by the \emph{Ordered Bell Number} $B_n$ which is between $n!$
and $2^{n-1}n!$~\cite[Seq.~A670]{Sloane}).
A brute-force solution, which always reaches the worst case, was also described.
This solution turned out useful in constructing global ranking functions and will also be
used for this purpose in this paper.

\subsection{Properties of stable systems}

\blem \label{lem:stablem}
Let $M = G_1G_2\dots G_{\ell}$ be a finite multipath of
a stable MCS. Suppose that there is a path in $M$ from $x[0,s]$ to $x[\ell,t]$. Then there is a thread
with such endpoints.  
\elem

The lemma works for down-paths (and down-threads) as well as for up-paths (and up-threads), which
can be seen (for more convenient argument) as paths in the transposed graph.

\bprf We consider the shortest path among the given endpoints.
It has to be a thread, for otherwise it has two consecutive arcs among the nodes of a single graph $G_i$.
Stability and consequence-closure of the graphs imply that these two arcs can always be replaced by
a single one, contradicting the choice of a shortest path. 
\eprf

\blem \label{lem:stablemm}
Let $M = G_1G_2\dots G_{\ell}$ be a finite multipath of
a stable MCS. Suppose that there is a path in $M$ from $x[t,i]$ to $x[t,j]$ for some $t,i,j$ with $i\ne j$.
 Then there is an arc with such endpoints.  
\elem

\noindent
The argument is very similar to the previous one and we omit a detailed proof.

\blem \label{lem:stable=satisfiable}
In a stable MCS, every finite multipath is satisfiable.
\elem

\noindent
The proof, again, is rather similar to the previous ones and is given in full in~\cite{BA:mcs}.

\section{Termination}
\label{sec:walks}

A central contribution of~\cite{leejonesbenamram01} was the definition of a 
``path based'' termination condition for systems of size-change graphs,
dubbed \emph{the SCT criterion}. 
The condition 
hinges on the presence of certain \emph{infinite paths} in infinite sequences
of size-change graphs (such sequences represent hypothetic execution histories).
 This criterion was generalized in~\cite{BA:mcs} to monotonicity constraints,
where the paths have become walks (possibly cyclic). It has also been shown
that assuming stability, the
condition reduces to the SCT criterion, which 
is easier to reason about as well as to test for.

In this section we prove that
for $\pi$-termination too, termination is captured by a path-based
criterion. We consider in particular the case of a stable system, but also the general case.
This yields a proof that $\pi$-termination is decidable in PSPACE.

\begin{ZvsWF}
The path-based criterion, and consequently the algorithm, enhance the criteria and algorithm of~\cite{BA:mcs}
in a way which seems, at least after the fact, very natural. The completeness proof was the challenging part as it differs significantly
from~\cite{BA:mcs}; it is inspired by~\cite{Codish-et-al:05}. This section also includes a discussion of the algorithm
proposed in~\cite{Codish-et-al:05}.
\end{ZvsWF}

\subsection{Some definitions}

\begin{fig0}{A multipath.}{t}{fig-multipath}
$$\xymatrix@R=20pt@C=40pt{
  x[0,1]\ar@<2ex>@/^10pt/@{->}[dd] & x[1,1] \ar@<2ex>@/^10pt/@{->}[dd] \ar@{->}[l] \ar@{->}[r] & x[2,1]\ar@<2ex>@/^10pt/@{->}[dd] & x[3,1] \ar@{->}[l]\\
  x[0,2] \ar@{->}[r]\ar@{->}[d]  & x[1,2] \ar@{->}[r]  & x[2,2]\ar@{->}[d]
  \ar@{->}[r]& x[3,2] \\
  x[0,3]\ar@/_4pt/@{-->}[r] & x[1,3]\ar@/_4pt/@{-->}[l]\ar@/_4pt/@{-->}[r]\ar@{->}[u]
  & x[2,3]\ar@/_4pt/@{-->}[l] \ar@/_4pt/@{-->}[r]& x[3,3]\ar@/_4pt/@{-->}[l]
}$$
\end{fig0}

\bdfn[multipath]
Let $\cal A$ be an $n$-variable MCS, and let 
$f_0\stackrel{G_1}{\to}f_1\stackrel{G_2}{\to}f_2\ldots$ be an MC-labeled path in the CFG
(either finite or infinite).
The {\em multipath} $M$ that corresponds to this path
is a (finite or infinite) graph with nodes $x[t,i]$, where $t$ ranges from 0
up to the length of the path (which we also refer to as the length of $M$),
and $1\le i\le n$. Its arcs are obtained by merging the following sets:
for all $t\ge 1$, $M$ includes the arcs of
$G_t$, with source variable $x_i$ renamed to $x[t-1,i]$ and target variable
$x_j'$ renamed to $x[t,j]$. 
\edfn

The multipath may be written concisely as $G_1G_2\dots$; if $M_1, M_2$ are
finite multipaths, $M_1$ corresponding to a CFG path that ends where $M_2$
begins, we denote by $M_1M_2$ the result of concatenating them
in the obvious way.  The notation $(G)^x$ represent a multipath made of $x$ copies of $G$.

Figure~\ref{fig-multipath} depicts multipath $G_1G_3G_1$, based on the MCs from Figure~\ref{fig:exintro}.

Clearly, a multipath can be interpreted as a conjunction of constraints
on a set of variables associated with its nodes. 
We consider assignments $\sigma$ to these variables, where the value
assigned to $x[t,i]$ are denoted $\sigma[t,i]$.
A multipath can be seen  as an execution trace of the abstract program, whereas
a satisfying assignment constitutes a (concrete) run of ${\cal T}_{\cal A}$. Conversely:
every run of ${\cal T}_{\cal A}$ constitutes a satisfying assignment to the corresponding multipath.

\bdfn
A path in the graph representation of an MC or a multipath is \emph{strict} if it includes a strict arc.
\edfn

\begin{obs}
A finite multipath is satisfiable if and only if it does not contain a strict cycle.
\end{obs}

We next define down-paths and up-paths. The definition of a down-path is just the standard definition
of path (which has 
already been used in this paper) but it is renamed in order to accommodate the notion of an up-path.

\bdfn 
A \emph{down-path} in a graph is a sequence $(v_0,e_1,v_1,e_2,v_2,\dots)$ where
for all $i$, $e_i$ is an arc from $v_{i-1}$ to $v_i$ (in the absence of
parallel arcs, it suffices to list the nodes).
An \emph{up-path} is a sequence
$(v_0,e_1,v_1,e_2,v_2,\dots)$ where
for all $i$, $e_i$ is an arc from $v_{i}$ to $v_{i-1}$.

The term path may be used generically to mean either a down-path or an up-path (such usage should be
clarified by context).
\edfn

Semantically, in an MC or a multipath, a down-path represents a descending chain of values,
whereas an up-path represents an ascending chain. Note also that an up-path is a down-path in the \emph{transposed} graph. 

\bdfn 
Let $M = G_1G_2\dots$ be a multipath.
A \emph{down-thread} in $M$ is a down-path that only includes arcs 
in a forward direction
($x[t,i] \to x[t+1,j]$).

An \emph{up-thread} in $M$
is an up-path that only includes arcs in a backward direction
($x[t,i] \gets x[t+1,j]$).

A \emph{thread} is either.
\edfn

\subsection{Combinatorial criteria for $\pi$-termination}

We formulate necessary and sufficient conditions for the termination of an MCS in terms of its multipaths.
The first criterion is called \emph{Condition S} as it assumes a stable MCS.
The second, \emph{Condition G}, works with any MCS.

\subsubsection{The Stable Case}
\label{sec:conS}

\begin{defi}
A stable MCS $\cal A$ satisfies \emph{Condition \textrm{S}} if in any infinite multipath $M$ (with variables
$x[t,i]$) there are an infinite down-thread
$(x[k, h_k])_{k=k_0,k_0+1,\dots}$ 
and an up-thread
$(x[k, l_k])_{k=k_0,k_0+1,\dots}$, such that all the constraints
$x[k, l_k]\le x[k,h_k]$ are present in $M$.
 In addition, at least one of the threads has
infinitely many strict arcs.
\end{defi}

We may later refer to the up-thread $(x[k, l_k])$ as the \emph{low thread}, while
the down-thread $(x[k, h_k])$ is the \emph{high thread}. This naming is meant
to stress the relation $x[k, l_k]\le x[k,h_k]$: the values assumed by the
variables on the low thread are always lower than those assumed on the high
thread. The fact that the values in the high thread descend, while those in the low thread
ascend, suggests that the process cannot go on forever, which is what we
want. A down-thread/up-thread pair satisfying this condition will be called \emph{an approaching pair}.
If the condition of infinitely many strict arcs is not guaranteed, we will use the term
\emph{a weakly approaching pair}.

\bthm
Condition \mbox{\rm S} is a sufficient condition for $\pi$-termination.
\ethm

\bprf
Suppose that Condition~S is satisfied and that $\cal A$ does not terminate. Thus,
some infinite multipath $M$
is satisfiable. The threads postulated in Condition~S imply that
the sequence of differences $\sigma[k,h_k] - \sigma[k,l_k]$ is 
infinitely descending, while consisting of non-negative integers.
 This is impossible,
so we conclude that $\cal A$ is $\pi$-terminating.
\eprf

Condition S is also a necessary condition for termination
of stable systems (i.e., the criterion is sound and complete), but we will prove this later.

Readers familiar with SCT will surely notice that if, for each flow-point $f$,
for every pair of variables such that $I_f\vdash x_i\le x_j$, we create a variable $x_{(i,j)}$ to represent $x_j-x_i$
(guaranteed to be non-negative) and we connect such variables with the obvious
size-change arcs (if $x_i\le x_l'$ and $x_j\ge x_h'$ then $x_{(i,j)}\ge x_{(l,h)}'$, etc),
Condition~S becomes equivalent to ordinary (well-founded) size-change termination in the new variables.
The SCT condition requires every infinite multipath to include an infinitely-descending thread
(a down-thread which is infinitely often strict). Our
approaching pairs correspond precisely to threads in the difference variables.

Thus, given a stable system, $\pi$-termination reduces to SCT, and this immediately provides us
with a decision algorithm. Nevertheless, we dedicate the next couple of subsections to new
decision algorithms, based directly on Condition~S. The motivation for doing so is threefold.
First, it is theoretically interesting to see how such a direct algorithm would work. Secondly,
the SCT algorithms are exponential in the number of variables. Creating difference variables in advance
squares the exponent, which is bad. Finally, we shall employ one version of the direct algorithm
in proving the \emph{completeness} of our termination criterion.

\subsubsection{The General Case}
\label{sec:conG}

Dropping the stability does not complicate the statement of the termination condition very much. Instead of
threads, we have to consider paths of any form (even such that repeat arcs). 

\begin{defi}
MCS $\cal A$ satisfies \emph{Condition \textrm{G}} if in any infinite multipath $M$ (with variables
denoted by $x[i,j]$)
 there is an up-path 
$(x[L_j, l_j])_{j=0,1,\dots}$ and a down-path
$(x[H_j, h_j])_{j=0,1,\dots}$ such that at least one of the paths has infinitely many strict arcs.
Moreover, for infinitely many values of $k$, there is a path
from $x[H_k, h_k]\le x[L_k,l_k]$. 
\end{defi}

\bthm
Condition \mbox{\rm G} is a
sound condition for $\ints$-termination.
\ethm

\bprf
Suppose that Condition~G is satisfied and that some infinite multipath $M$
is satisfiable. Then the paths described in Condition~G exist and imply that
the sequence of differences $x[H_j,h_j]-x[L_j,l_j]$ is 
infinitely descending, while consisting of non-negative integers
(note that if $x[L_k, l_k]\le x[H_k,h_k]$, then also for all $j\le k$, $x[L_j, l_j]\le x[H_j,h_j]$).
 This is impossible,
hence no infinite multipath is satisfiable, or equivalently, the MCS
$\ints$-terminates.
\eprf

\bthm \label{thm:G=S}
Conditions \mbox{\rm G} and \mbox{\rm S} are equivalent for a
stable system.
\ethm
\bprf
Condition~S is a special caes of Condition~G, so all we have to prove is that if
Condition~G is satisified, so is Condition~S.  Suppose, then, that Condition~G
holds, and we have the paths $(x[L_j, l_j])_{j=0,1,\dots}$ and 
$(x[H_j, h_j])_{j=0,1,\dots}$, as above. 
It is easy to see that since for infinitely many values of $k$, there is a path
from $x[H_k, h_k]$ to $x[L_k,l_k]$, it is also true that for \emph{any} $k$ and $j$, there is a path
from $x[H_k, h_k]$ to $x[L_j,l_j]$ (the path may use parts of the up-path and down-path
themselves).

 To each path, we can apply Lemma~\ref{lem:stablem} to obtain a \emph{thread} whose nodes
are a subset of the nodes of the path; an up-thread $(x[t, l_{j_t}])_{t\ge t_0}$
and a down-thread $(x[t, h_{k_t}])_{t\ge t_0}$. At least one of the threads 
is infinitely often strict.

From the previous observation, we know
that there is, for every $t$, a path from $x[t, h_{k_t}]$ to $x[t, l_{j_t}]$.  Applying 
Lemma~\ref{lem:stablemm}, we cam show that the
relation $x[t, h_{k_t}] \ge x[t, l_{j_t}]$ must be included in the invariant for the 
corresponding flow-point.
\eprf

Condition G is also a complete criterion for termination; the proof is based on stabilizing the system and the equivalence
of Conditions~G with Condition~S in the stable one.  The details are omitted as they are tedious and give no new insight.

\subsection{A Closure Algorithm for Stable Systems}
\label{sec:intClosure}

We next present an algorithm to decide termination of stable Integer MCS by computing a composition-closure
and applying a certain test to the elements of the closure set.
This algorithm is based on the Closure Algorithm for the well-founded model~\cite{LS:97,CT:99,DLSS:2001,leejonesbenamram01,BA:mcs},
and is theoretically related to the Disjunctive Well-Foundedness principle~\cite{HJP:2010}.
An algorithm \emph{sans} stability is presented later for completeness; but our analysis of the algorithms and their correctness
makes substantial use of stability (note the previous paragraph).

The following definitions are essentially from~\cite{BA:mcs}:

\bdfn[composition] \label{def-composition}
The {\em composition} of MC $G_1:
f\to g$ with $G_2: g\to h$, written $G_1;G_2$, is a MC with source
{$f$} and target {$h$}, which includes 
all the constraints among $s,s'$ implied by
$\exists s'' : \trans{G_1}{s}{s''}\land \trans{G_2}{s''}{s'}$.
\edfn

Composition is similar to logical closure and can be implemented by a DFS-based
algorithm in $O(n^3)$ time. 

\bdfn[collapse]
For a finite multipath $M = G_1\dots G_\ell$,
Let $\overline M = \overline{G_1;\cdots;G_\ell}$. This is called the
\emph{collapse} of $M$ (if $\ell =1$, it is just the consequence-closure of $G_1$).
\edfn

Applying composition together with logical closure, we can easily compute of $\overline M$ for a given $M$.

\bdfn
Given an MCS $\cal A$,  its closure set $\clos{\cal A}$ is
$$\{\overline M \mid M \mbox{ is a satisfiable finite $\cal A$-multipath}\}. $$
\edfn

\noindent
The closure set can be computed by a routine least-fixed-point procedure, as in~\cite{BA:cav09}.

\bdfn[cyclic MC]
We say that a MC $G$
is \emph{cyclic} if its source and target flow-points are equal.
This is equivalent to stating that $GG$ is a valid multipath.
\edfn

\bdfn[circular variant]
For a cyclic $G$, 
the {\em circular variant\/} $G^\cv$ is a  weighted graph
obtained by adding, for every variable $x_i$, an edge
$x_i  \edge x_i'$. This edge is treated as a pair of no-change arcs,
but is distinguished from edges already present in $G$. These
additional edges are called \emph{shortcut edges}.
\edfn

\bdfn[types of cycles]
Let $G$ be a cyclic MC.
A  cycle in $G^\cv$ is a  path commencing and ending at
 the same node.  It is a \emph{forward cycle}
if it traverses shortcut edges only in the backward
direction (from $x_i'$ to $x_i$)%
\footnote{this naming may seem strange, but will will later
see that such a cycle is ``unwound'' into a forward-going thread in the
multipath $(G)^\omega$.}.
It is a \emph{backward} cycle if it
traverses shortcut edges only in the forward direction.
\edfn

\bdfn[Local $\pi$-Termination Test for Stable Systems]
For a cyclic $G$, 
we say that $G$ passes the \emph{Local $\pi$-Termination Test for Stable Systems}, or LTTS, if $G^\cv$
includes a forward 
cycle $F$ and a backward cycle $B$,
at least one of which is strict, and an arc from a node
of $F$ to a node of $B$.
\edfn

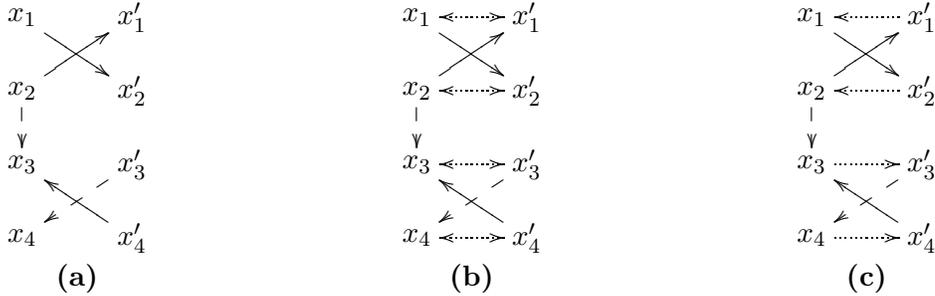
\begin{figure}[t]
\begin{centering}
\begin{tabular*}{0.9\textwidth}{@{\extracolsep{\fill}}ccc@{\extracolsep{0pt}}c}
$\xymatrix@R=10pt{
  x_1\ar@{->}[dr] & x_1' \\
  x_2\ar@{->}[ur]\ar@{-->}[d]  & x_2' \\
  x_3        & x_3'\ar@{-->}[ld] \\
  x_4        & x_4'\ar@{->}[lu]
}$ &
$\xymatrix@R=10pt{
  x_1\ar@{->}[dr] & x_1'\ar@{<.>}[l] \\
  x_2\ar@{->}[ur]\ar@{-->}[d]  & x_2'\ar@{<.>}[l] \\
  x_3\ar@{<.>}[r] & x_3'\ar@{-->}[ld] \\
  x_4\ar@{<.>}[r] & x_4'\ar@{->}[lu]
}$ &
$\xymatrix@R=10pt{
 x_1\ar@{->}[dr] & x_1'\ar@{.>}[l] \\
  x_2\ar@{->}[ur]\ar@{-->}[d]  & x_2'\ar@{.>}[l] \\
  x_3\ar@{.>}[r] & x_3'\ar@{-->}[dl] \\
  x_4\ar@{.>}[r] & x_4'\ar@{->}[ul]
}$ \\ 
 {\bf (a)} & {\bf (b)} & {\bf (c)} \\
\end{tabular*}\par

\end{centering}

\caption{(a) An MC $G$, (b) its circular variant $G^\circ$; (c) the directions of
shortcut edges are set  to highlight the two cycles required by the LTTS. }
\label{fig:ltts}
\end{figure}

This test resembles Sagiv's local test for the well-founded case~\cite{LS:97}, and likewise
can be implemented as a DFS-based algorithm in linear time, on which we will not 
elaborate.
Figure~\ref{fig:ltts} illustrates the test.

\begin{algo} (The Closure Algorithm) \label{alg-closure}
\end{algo}
\be
\item Build $\clos{\cal A}$.
\item
For each  cyclic $G$ in $\clos{\cal A}$, apply the LTTS. \\
Pronounce failure (non-termination) if a graph that fails the
test is found.

\item 
 If the previous step has completed, the MCS terminates.
\ee

\bthm[closure algorithm---soundness] \label{thm:LTTSimpliesCondS}
Let $\cal A$ be a stable MCS.
If every cyclic MC in $\clos{\cal A}$ passes the Local $\pi$-Termination Test for Stable Systems,
$\cal A$ satisfies Condition~\mbox{\rm S}.
\ethm

The proof will use the next lemma.

\begin{lem} \label{lem:segments}
Consider an infinite multipath $M$ of a stable system, represented as
the concatenation of finite segments $M_1M_2\dots$, 
and let ${M}' =
(\overline M_1)(\overline M_2)\dots$. Then 
${M}'$ satisfies Condition~\mbox{\rm S} if and only if $M$ does.
\end{lem}

\bprf
Suppose first that $M'$ satisfies Condition~S.  The down-thread in $M'$ can be mapped back
to $M$, where it consists of variables $x[k_i, h_i]$, such that $k_i$ is the length of
$M_1\dots M_i$. The arc $x[i,h_i]\to x[i+1,h_{i+1}]$ in $M'$ is actually an arc of $\overline M_{i+1}$,
and reflects (by the definition of composition) a path in $M_{i+1}$. By Lemma~\ref{lem:stablem},
this path can be assumed to be a thread. We thus obtain an infinite down-thread that passes through
the variables $x[k_i, h_i]$. Similarly, we obtain an infinite up-thread through the variables $x[k_i, l_i]$.
Condition~S also implies that the constraint $x[k_i, h_i] \ge x[k_i, l_i]$ is present in $M$
for every $i$. For $k_{i-1} < k < k_{i}$, we have a down-thread from some $x[k,h]$ to $x[k_i,h_i]$,
and an up-thread from some $x[k,l]$ to $x[k_i,l_i]$, so we have a path from $x[k,h]$ to $x[k,l]$. 
By Lemma~\ref{lem:stablemm},  $x[k,h]$ and $x[k,l]$ must be explicitly related in $M$.

For the converse implication, we assume that $M$ satisfies Condition~S and we are to prove it
for $M'$: this is symmetric, but simpler than the above argument, and so is left to the interested reader.
\eprf

\bprf (of Theorem~\ref{thm:LTTSimpliesCondS})
We suppose that
 every cyclic MC in $\clos{\cal A}$ passes the LTTS.
Let ${\cal M}=G_1G_2\ldots$ be any infinite $\cal A$-multipath.

Consider the set of positive integers,
and label each pair $(t,t')$ , where $t<t'$, by
\[
G = G_{t}; G_{{t+1}}; \cdots G_{{t'-1}}
\]
which is included in $\clos{\cal A}$, since this multipath is satisfiable (Lemma~\ref{lem:stable=satisfiable}).
By Ramsey's theorem (in its infinite version), there is an infinite set of
positive integers, $I$, such that all pairs $(t,t')$ with $t,t'\in I$
carry the same label $G_I$.

Thus for any $t,t'\in I$ with $t<t'$,
$G_{t}; G_{{t+1}}; \cdots G_{{t'-1}} = G_I$.
By Lemma~\ref{lem:segments}, it now suffices to show that multipath
$(G_I)^\omega$ (infinite sequence of $G_I$'s) satisfies Condition~S.

By assumption, $G_I$ passes the LTTS. Let $F$ be the forward cycle.
Due to consequence-closure, we can shrink each
segment of $F$ that consists of ordinary arcs (not shortcuts) into a single arc, so that without 
loss of generatlity we may assume that the
cycle alternates source (unprimed) with target (primed) variables.
We can thus choose indices $h_1,h_2,\dots$ such that the nodes of the cycle are
$x_{h_0}$, $x_{h_1}'$,  $x_{h_1}$, $x_{h_2}'$ and so on up to $x_{h_{s}}'$ for some
$s>0$, with $h_{s} = h_0$.
We can do the same for the backward cycle $B$, denoting the variables by
$x_{l_j}$ and $x_{l_j}'$, for $0\le j< \hat s$, where $ \hat s$ is the length of that cycle.
By LTTS, there is also an arc from a node of $F$ to a node of $B$; suppose that they are both
source nodes; then, without loss of generality (a cycle can be begun at any point),
we can assume that $G\vdash x_{h_0}\ge x_{l_0}$.
In fact, there is no loss of generality in assuming that the related nodes are source nodes, either;
suppose, for example, that we have $G\vdash x_{h_1}'\ge x_{l_0}$; then 
$G\vdash x_{h_0}\ge x_{l_0}$ is implied. Other situations can be handled similarly.

To show that Condition~S is satisfied, we map the cycles onto infinite threads in $(G_I)^\omega$.
The $k$th node of the down-thread is $x[k, h_{k \bmod s}]$.
The $k$th node of the up-thread is $x[k, l_{(\hat s - k) \bmod \hat s}]$ (indexing the backward cycle
with $\hat s - k$ rather than $k$ yields an up-thread).
Either the infinite down-thread or the infinite up-thread (or both) is infinitely often strict.

According to our assumptions, for all $q$ we have the constraint $x[qs\hat s, h_0] \ge x[qs\hat s, l_0]$.
This implies $x[k, h_{k \bmod s}] \ge x[k, l_{k \bmod \hat s}]$ since we have the following path:
traversing part of the up-thread from
$x[k, h_{k \bmod s}]$ to $x[qs\hat s, h_0]$ for some $q$ such that $qs\hat s \ge k$,
then using
$x[qs\hat s, h_0] \ge x[qs\hat s, l_0]$ which we have by assumption, and then traversing
the down-thread in the reverse direction from  $x[qs\hat s, l_0]$ to $x[k, l_{k \bmod \hat s}]$.
By Lemma~\ref{lem:stablemm}, the deduced constraint must explicitly appear in in the multipath.
 
We conclude that Condition~S is satisfied by $(G_I)^\omega$.
\eprf

\subsection{The Role of Idempotence}

We call $G$ \emph{idempotent} if $G$ is cyclic and $G;G=G$.
Following~\cite{Codish-et-al:05} (and similar results in~\cite{DLSS:2001,leejonesbenamram01}), we claim:
\be
\item
In the closure algorithm, it suffices to test only idempotent members of $\clos{\cal A}$.
\item
The local test becomes simpler with idempotent graphs.
\ee

The first claim is easy to justify by studying the argument in the soundness proof:
the graph $G_I$ whose existence is established there is clearly idempotent.
For the second, let us describe the simple test and justify it.

\bdfn[Local $\pi$-Termination Test for Idempotent MCs]  \label{def:LTT1}
We say that $G$ passes the \emph{Local $\pi$-Termination Test for Idempotent MCs}, or LTT1,
if it is idempotent, and for some $1\le l,h\le n$, 
$G\vdash x_l\le x_h \land x_l\le x_l' \land x_h\ge x_h'$, where at least one of the last two
inequalities is strict.
\edfn

\blem \label{lem:LTT1}
Let $\cal A$ be a stable MCS and $G$ an idempotent, cyclic MC in $\clos{\cal A}$.
Then $G$ passes the LTTS if and only if it passes the LTT1.
\elem

\bprf
One direction of this equivalence is easy: if $G$ passes the LTT1, it has the cycles $x_h\to x'_h\to x_h$
and $x_l\to x'_l\to x_l$, that satisfy the LTTS.

For the other direction, 
suppose that $G$ passes the LTTS.  
Repeating the analysis in the proof of Theorem~\ref{thm:LTTSimpliesCondS}, and re-using the notation,
we let $2s$ be the length of the forward cycle (assuming it alternates shortcut and ordinary arcs) and $2\hat s$ the
length of backward cycle (under a similar assumption), and 
consider the multipath ${\cal M}=(G)^{s\hat s}$.   It has a down-thread from $x[0,h_0]$ to $x[s\hat s,h_0]$, so 
$\overline{(G)^{s\hat s}}\vdash x_h\ge x_h'$. But by idempotence, $\overline{(G)^{s\hat s}} = G$.
Similarly, we deduce that $G\vdash x_l\le x_l'$,  that one of these relations is strict, and that
$G\vdash x_l\le x_h$. Therefore, $G$ passes the LTT1.
\eprf

We obtain the following version of the closure algorithm:
\begin{algo} (Closure Algorithm with Idempotence) \label{alg-closure-I}
\end{algo}
\be
\item Build $\clos{\cal A}$.
\item
For each  cyclic $G$ in $\clos{\cal A}$, if it is idempotent apply the LTT1
to $G$. Pronounce failure (non-termination) if a graph that fails the
test is found.
\item 
 If the previous step has completed, the MCS terminates.\smallskip
\ee

\noindent Though this algorithm makes fewer local tests, it does not seem to be practically better than
the previous, since the local test is quite efficient. On the contrary, applying the test to every
graph in the closure allows for early discovery of failure. More importantly, it allows for reducing
the size of the set by subsumption (consider~\cite{BL:2006,FV:09} and others).

\subsection{Complexity}

The closure algorithm, regardless of minor savings based on idempotency or subsumption, has an
exponential worst-case for time and space. This follows from the fact that the closure set can be exponential.
An easy upper bound is this: for any pair of flow points $f,g$, an MC relates the $2n$ variables in the source
and target states. The number of such MCs is easily bounded by $3^{4n^2}$ as an MC is completely
specified by choosing either $\le$, $<$ or ``nothing'' for every ordered pair of variables.
The complexity of the whole data structure, and the time of the algorithm, are thus
$\text{poly}(m)\cdot 2^{O(n^2)}$ where $m$ is the number of flow-points%
\footnote{There are instances of size $O(n)$ that actually generate a closure of 
$2^{\Theta(n^2)}$ MCs. The exponent is thus not an overestimate.}.

In Section~\ref{sec:elaborate}, a polynomial-space algorithm will be described, just to make the theoretical 
classification of the problem in PSPACE. Practically, the polynomial-space version is not attractive because
its running time gets worse. On the other hand, in Section~\ref{sec:walks} we show that the exponent in the
running time can be brought down to $O(n\log n)$. But this, too, seems to be an improvement only in a 
highly theoretical sense.

\subsection{Completeness}

This section proves the completeness of Condition~S, as well as the closure algorithm,
by proving that if an idempotent MC in a stable system does not satisfy the LTTS, the 
system fails to terminate.

To show that the system does not terminate, we exhibit a satisfiable infinite multipath. 

\blem \label{lem:extend}
Let $M$ be a finite, cyclic, satisfiable
 multipath and $G = \overline M$. If $(G)^\omega$ is satisfiable,
so is $(M)^\omega$.
\elem

\bprf
Suppose that  $(G)^\omega$ is satisfiable. That is, a satisfying assignment $\sigma$ exists
such that $\sigma[t,i]$ is the value of $x[t,i]$.  Consider $(M)^\omega$. For distinction, let
us denote its variables by $x'[\_,\_]$.
Let $\ell$ be the length of $M$; the variables $x'[\ell t, i]$ represent the borders of the
copies of $M$, and correspond to the source and target variables of $G$.
Recall that $i$ varies between $1$ and $n$, the number of variables.
Set $\sigma'[\ell t, i] = n(\ell+1)\sigma[t,i]$. We claim that $\sigma'$ can be extended
to a satisfying assignment for $(M)^\omega$.  The construction is illustrated in Figure~\ref{fig:assignment}.

When considering the assignment to a particular copy of $M$, it is possible to ignore all
others, since all nodes on its boundary are already assigned. Thus, suppose that we are
given a copy of $M$, say the $p\,$th copy,
 with an assignment to its boundary nodes given by $\sigma'[p\ell,\_]$
and $\sigma'[(p+1)\ell,\_]$. Consider $M$ as a weighted graph;
the graph has no negative-weight cycles, or $M$ would be unsatisfiable. 

Let $\mu$ be the biggest value assumed by $\sigma'$ over the boundary nodes of $M$.
Add an auxiliary node $z$ to $M$ and
arcs from $z$ to all $M$'s nodes, weighted as follows: an arc to a boundary node $x'[(p+j)\ell,i]$
is weighted by  $\sigma'[(p+j)\ell,i]$. All other arcs have weight $\mu + n(\ell+1)$
(note that $n(\ell+1)$ bounds the length of any simple path in $M$).
Let $\delta_z(j,i)$ be the weight of the
lightest path from $z$ to $x'[j,i]$. This is well-defined, due
to the absence of negative cycles. 

If there is an arc of weight $w$ from $x'[j,i]$ to $x'[j',i']$, then $\delta_z(j',i') \le \delta_z(j,i)+w$.
Recalling that non-strict arcs have a weight of 0 and strict arcs, of $-1$,
it is easy to see that assigning $\delta_z(j,i)$ to $x'[j,i]$
is a satisfying assignment for $M$.

It rests to show that this assignment agrees with $\sigma'$. To this end, let $x[(p+j)\ell,i]$ be a 
boundary node.
 By construction, $\delta_z((p+j)\ell, i) \le \sigma'[(p+j)\ell, i]$.
A strict inequality can only arise if there is a path from $z$ to $x'[(p+j)\ell,i]$,
lighter than the immediate arc.  It can be assumed to be a simple path;
it begins with an arc $z\to x'[r,s]$ and then rests within $M$. Let $P$ be the segment
within $M$; since it is simple,
 its weight $w$ is bigger than $-n(\ell+1)$. Using this, one can easily eliminate the possibility
that the path begins with an arc $z\to x'[r,s]$ where $x'[r,s]$ is not a boundary node.
Suppose, then, that $x'[r,s]$ \emph{is}
 a boundary node, and can be written as $x'[(p+k)\ell, s]$.
Then, we have
\begin{equation} \label{eq:s+w}
\sigma'[(p+k)\ell, s] + w < \sigma'[(p+j)\ell, i] \,.
\end{equation}
We now distinguish two cases. If $\sigma[p+k,s] > \sigma[p+j,i]$, we get (multiplying by $n(\ell+1)$):
\begin{equation}
\sigma'[(p+k)\ell, s] \ge n(\ell+1) + \sigma'[(p+j)\ell, i]
\end{equation}
contradicting Inequality~(\ref{eq:s+w}).

If $\sigma[p+k,s] = \sigma[p+j,i]$, there can be no strict path from $x[p+k,s]$ to $x[p+j,i]$, and hence no
strict path from $x'[(p+k)\ell, s]$ to $x'[(p+j)\ell, i]$. Thus $w\ge 0$; and again a contradiction ensues.

If $\sigma[p+k,s] < \sigma[p+j,i]$, there can be no path at all from $x[p+k,s]$ to $x[p+j,i]$, and hence no
path from $x'[(p+k)\ell, s]$ to $x'[(p+j)\ell, i]$. 

We conclude that  $\delta_z((p+j)\ell, i) = \sigma'[(p+j)\ell, i]$, 
so we have a satisfying assignment that extends $\sigma'$, as desired.
\eprf

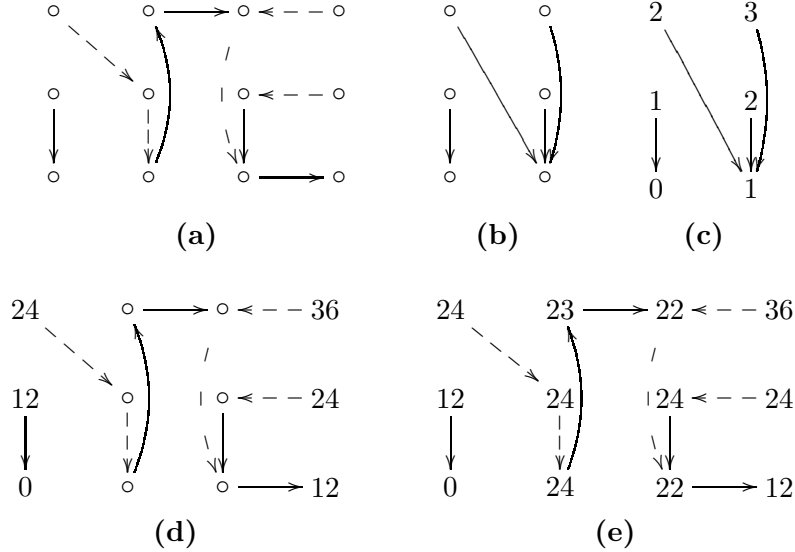
\begin{fig0}{An illustration for the proof of Lemma 4.23:
(a) A multipath $M$; (b) $G = \overline M$; (c) $G$ with an assignment $\sigma$;
(d) $M$ with its boundary values set (note that $n(\ell+1) = 3\cdot 4 = 12$);
(e) the assignment to $M$ completed by a lightest-path computation.
}{t}
{fig:assignment}
\begin{centering}
\setlength{\extrarowheight}{1ex}
\begin{tabular}{@{\extracolsep{20pt}}ccc}
$\xymatrix@R=20pt{
  \circ\ar@{-->}[dr]& \circ\ar@{->}[r]&\circ\ar@/_8pt/@{-->}[dd] &\circ\ar@{-->}[l]  \\
  \circ\ar@{->}[d] & \circ\ar@{-->}[d] &\circ\ar@{->}[d]&\circ\ar@{-->}[l] \\
  \circ     & \circ\ar@/_8pt/@{->}[uu] &\circ\ar@{->}[r] &\circ 
}$ &
$\xymatrix@R=20pt{
  \circ\ar@{->}[ddr]& \circ\ar@/^6pt/@{->}[dd]  \\
  \circ\ar@{->}[d] & \circ\ar@{->}[d] \\
  \circ     & \circ 
}$ &
$\xymatrix@R=20pt{
  2\ar@{->}[ddr]& 3\ar@/^6pt/@{->}[dd]  \\
  1\ar@{->}[d] & 2\ar@{->}[d] \\
  0     & 1
}$ 
 \\
\textbf{(a)} & \textbf{(b)}& \textbf{(c)}  \\
\end{tabular}\par\bigskip
\begin{tabular}{@{\extracolsep{20pt}}cc}
$\xymatrix@R=20pt{
  24\ar@{-->}[dr]& \circ\ar@{->}[r]&\circ\ar@/_8pt/@{-->}[dd] &36\ar@{-->}[l] \\
  12\ar@{->}[d] & \circ\ar@{-->}[d] &\circ\ar@{->}[d]&24\ar@{-->}[l]  \\
    0     & \circ\ar@/_8pt/@{->}[uu] &\circ\ar@{->}[r] &12
}$ &
$\xymatrix@R=20pt{
  24\ar@{-->}[dr]& 23\ar@{->}[r]& 22\ar@/_8pt/@{-->}[dd] &36\ar@{-->}[l] \\
  12\ar@{->}[d] & 24\ar@{-->}[d] & 24\ar@{->}[d]&24\ar@{-->}[l]  \\
    0     & 24\ar@/_8pt/@{->}[uu] & 22\ar@{->}[r] &12
}$
 \\
 \textbf{(d)} & \textbf{(e)} \\
\end{tabular}\par
\end{centering}
\end{fig0}

\blem \label{lem:LTTScomplete}
Let $\cal A$ be a stable MCS and $G\in \clos{\cal A}$ be an idempotent MC that does not
pass the Local $\pi$-Termination Test  for Stable Systems. Then $\cal A$ has a satisfiable, infinite multipath.
\elem

\bprf 
We will show that $(G)^\omega$ is a satisfiable multipath.
Then, the conclusion will follow by Lemma~\ref{lem:extend}.

The first step is to add a new variable $x_0$ (``the zero line'') with constraint $x_0=x_0'$.
Next, we make sure that every other variable is
related to $x_0$. We do this as follows:
\be
\item
For each $x_i$ such that $G\vdash x_i < x_i'$, we add $x_i > x_0$ and $x_i' > x_0'$.
\item
For each $x_i$ such that $G\vdash x_i > x_i'$, we add $x_i < x_0$ and $x_i' < x_0'$.
\item
If there is a variable $x_i$ 
which is still unrelated to $x_0$,
we add $x_i > x_0$ and $x_i' > x_0'$. This is repeated until all variables
are related to $x_0$ or $x_0'$.
\ee
It is not hard to verify that $G$ remains
idempotent throughout this process.
$G$ also remains satisfiable, since
the above additions are non-contradictory: It is easily seen that in the first stage, the additions 
cannot be contradictory---they cannot form a directed cycle; the additions in the 
second stage cannot be contradictory among themselves, and they cannot form 
a contradiction with the additions of the first stage, because, by assumption, we cannot have
$x_i < x_i'$ in $G$ together with $x_j > x_j'$ and $x_j \ge x_i$ (or $x_j' > x_i'$).
In the third stage, it is obvious that no contradictions can arise.

Let $\widehat G$ denote the extended MC; we will show that $(\widehat G)^\omega$ is
satisfiable.
Let $\cal V$ be the set of variables of $(\widehat G)^\omega$.
We define an assignment $\sigma:{\cal V}\to \ints$ as follows:
\bi
\item
For all $t$, $\sigma(x[t,0]) = 0$.  We shall refer to these variables collectively as $\mathbf{0}$
(note that they are all related by equalities so if a variable is related to one of them, it is related to all).
\item
For all $v\in{\cal V}$ such that there is a directed path from $v$ to $\mathbf{0}$,
let $\sigma(v)$ be the maximum number of strict arcs on such a path (note that
only simple paths need be considered, since there are no strict cycles).
\item
For all $v\in {\cal V}$ such that there is a directed path from $\mathbf{0}$ to $v$,
let $\sigma(v)$ be the negation of the maximum number of strict arcs on such a path.
\ei
Note that since every variable in $G$ is related to $x_0$ (or $x_0'$), we have covered all
cases. Why are the assignments well-defined? The only problem to worry about is the
existence of a node $v$ such that its distance from $\mathbf{0}$ is unbounded.

Suppose that such $v$ exists and that there is a path from $v$ to $\mathbf{0}$ (the other
case, a path from $\mathbf{0}$ to $v$, is symmetric).
Thus, there are infinitely many paths ${P}_i$ from $v$ to $\mathbf{0}$, such
that ${P}_i$ has at least $i$ strict arcs. For each such path, 
we can assume that only one $\mathbf{0}$ appears on the path; otherwise we can cut
the path at the first occurrence (there can be no strict arcs among occurrences of $\mathbf{0}$).

The pigeon-hole principle shows that for sufficiently large $i$, ${P}_i$ must
visit a certain state-variable $x_j$ twice, say as $x[t,j]$ first and $x[t',j]$ later,
with a strict arc in-between. Moreover, by choosing $i$ big enough, 
we can enforce the condition $t'>t$.
Then, by idempotence of $\widehat G$, we have 
$\widehat G \vdash x_j > x_j'$. But then, $x_j < x_0$ is in $\widehat G$; the rest of
${P}_i$ implies that there is a path from $x[t',j]$ to $\mathbf{0}$, so we also have
$G\vdash x_j' \ge x_0$, a contradiction.

We conclude that our assignment $\sigma$ is well defined; and the fact that it satisfies
all constraints is easy to prove by its construction. Since it satisfies $(\widehat G)^\omega$,
it also satisfies $(G)^\omega$.
\eprf

\begin{thm} \label{thm:Scomplete}
If a stable MCS does not satisfy
Condition \mbox{\rm S}, it is not $\pi$-terminating.
\end{thm}

\bprf
Let $\cal A$ be such a system; suppose that every idempotent MC in $\clos{\cal A}$ passes the LTT1.
The Condition~S is satisfied, contradictory to assumption. So there has to be an idempotent MC which
fails the LTTS, and by the last lemma, a non-terminating run exists.
\eprf

\begin{cor}
Condition \mbox{\rm S} is a
equivalent to 
$\pi$-termination of a stable MCS, and
the closure algorithm is a sound and complete decision procedure.
\end{cor}

\subsection{Codish, Lagoon and Stuckey's Algorithm}

Codish, Lagoon and Stuckey~\cite{Codish-et-al:05} also gave (albeit implicitly)
an algorithm to decide
 termination of an integer MCS, under the assumption that the data represented are non-negative integers
(so termination follows either by descent towards zero, or by approaching variables).
 The algorithm is closely related to Algorithms~\ref{alg-closure} and~\ref{alg-closure-I}. 

\bdfn[balanced constraint, \cite{Codish-et-al:05}]
A cyclic monotonicity constraint $G$ is \emph{balanced} if 
$G\vdash x_i \rhd x_j ~\Leftrightarrow~ G\models x'_i \rhd x'_j$ (where $\rhd$ is $>$ or $\ge$).
The \emph{balanced extension} $G_B$ of $G$
is the weakest monotonicity constraint which is at least as strong as $G$
and is balanced. 
\edfn

To compute the balanced extension, Codish, Lagoon and Stuckey suggest a fixed-point computation:
Define
$
\textit{bal}(G) = 
     G
     \wedge 
     \{ x_i \rhd x_j ~\mid~ G \vdash x'_i \rhd x'_j\}
     \wedge  
     \{ x'_i \rhd x'_j ~\mid~ G \vdash x_i \rhd x_j\}
$.
They observe that there always is a $p$ such that $\textit{bal}^{(p)}(G)$ is balanced and equals $G_B$
(it is then a fixed point of $\textit{bal}$).
Since there are at most $4n(n-1)$ constraints that can be added by $bal$,
$p\le 4n(n-1)$ (according to~\cite{Codish-et-al:05}, there are tighter bounds).
Hence, $G_B$ can be computed from $G$ in polynomial time.

\begin{algo} (CLS Algorithm) \label{alg-cls}
\end{algo}
Input: an MCS $\cal A$, \emph{not necessarily stable}.
\be
\item Build $\clos{\cal A}$.
\item
For each  cyclic $G$ in $\clos{\cal A}$, apply  the LTTS to $G_B$. \\
Pronounce failure (non-termination) if a graph that fails the test is found.
\item 
 If the previous step has completed, the MCS terminates.\smallskip
\ee

The CLS algorithm relies on
the balancing procedure to compensate for lack of stability in $\cal A$.  In fact, the significance of $G_B$ can be related to
stability.
Consider the MCS $\{G\}$, consisting of a single cyclic MC.  Let $S\{G\}$ be the smallest stable system bisimulating $\{G\}$
It is not hard to prove
that in $S\{G\}$ there is a single cyclic graph, namely $G_B$.

Soundness and completeness of Algorithm~\ref{alg-cls} are quite similar to those given for 
 Algorithms~\ref{alg-closure} and~\ref{alg-closure-I}. 
Codish et al.~give a partial completeness proof, specifically they prove that if $G_B$ fails the test
than $\{G\}$ does not terminate. The piece missing for proving non-termination of $\cal A$  is precisely Lemma~\ref{lem:extend}.

\subsection{A General Closure Algorithm}
\label{sec:gClosure}

We next present an algorithm which computes the composition-closure 
and applies a local test, which does not rely on stability. In essence, it corresponds to
Algorithm~\ref{alg-closure} in the way that Condition~G (Section~\ref{sec:conG}) corresponds to Condition~S;
as Condition~G refers to general paths instead of (up or down) threads, so will this test refer to cycles in the graph
that may have a more complicated form than those used by Algorithm~\ref{alg-closure}.
Since this is a strict generalization and does not conflict with the prior definitions, we use the same terminology for cycles,
redefined as follows.

\bdfn[types of cycles, generalized]
Let $G$ be a cyclic MC.
A  cycle in $G^\cv$ is a \emph{forward cycle}
if it traverses shortcut edges more often in the backward
direction (from $x_i'$ to $x_i$) than it does in the forward direction;
a \emph{backward} cycle, if it
traverses shortcut edges more often in the forward direction;
and a \emph{balanced cycle}
if it traverses shortcut edges equally often in both
directions.
\edfn

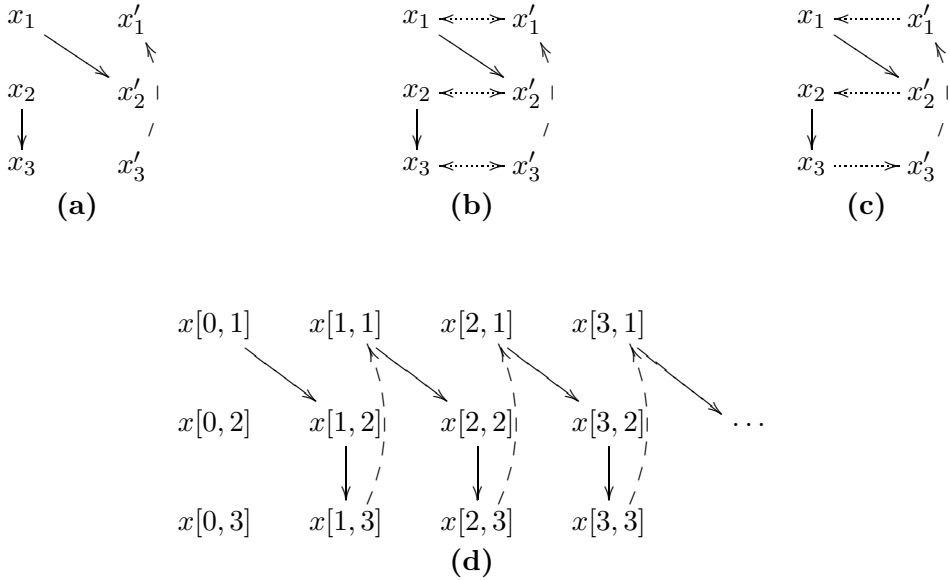
\begin{figure}[t]
\begin{centering}
\begin{tabular*}{0.9\textwidth}{@{\extracolsep{\fill}}ccc@{\extracolsep{0pt}}c}
$\xymatrix@R=10pt{
  x_1\ar@{->}[dr] & x_1' \\
  x_2\ar@{->}[d]  & x_2' \\
  x_3        & x_3'\ar@/_10pt/@{-->}[uu]
}$ &
$\xymatrix@R=10pt{
  x_1\ar@{->}[dr] & x_1'\ar@{<.>}[l] \\
  x_2\ar@{->}[d]  & x_2'\ar@{<.>}[l] \\
  x_3\ar@{<.>}[r] & x_3'\ar@/_10pt/@{-->}[uu]
}$ &
$\xymatrix@R=10pt{
 x_1\ar@{->}[dr] & x_1'\ar@{.>}[l] \\
  x_2\ar@{->}[d]  & x_2'\ar@{.>}[l] \\
  x_3\ar@{.>}[r] & x_3'\ar@/_10pt/@{-->}[uu]
}$ \\ 
 {\bf (a)} & {\bf (b)} & {\bf (c)} \\
\end{tabular*}\par
\end{centering}
\vspace{1cm}
\begin{centering}
\begin{tabular}{c}
$\xymatrix@C=15pt@R=20pt{
 x[0,1]\ar@{->}[dr] & x[1,1]\ar@{->}[dr] & x[2,1]\ar@{->}[dr] & x[3,1]\ar@{->}[dr]&\\
  x[0,2] & x[1,2]\ar@{->}[d] & x[2,2]\ar@{->}[d] & x[3,2]\ar@{->}[d] &\quad \dots\\
  x[0,3] & x[1,3]\ar@<-1ex>@/_10pt/@{-->}[uu]& x[2,3]\ar@<-1ex>@/_10pt/@{-->}[uu]& x[3,3]\ar@<-1ex>@/_10pt/@{-->}[uu] & \\
}$\\  {\bf (d)} \\
\end{tabular}\par
\end{centering}

\caption{(a) An MC $G$, (b) its circular variant $G^\circ$; (c) the directions of
shortcut edges are set  to form
a forward descending cycle, (d)
a walk in a prefix of $G^\omega$, corresponding to the cycle. 
The notation $x^t_i$ is a shorthand for $x[t,i]$.
(Example after
Codish, Lagoon and Stuckey)}
\label{fig:ltt}
\end{figure}

Figure~\ref{fig:ltt} (a--c) illustrates a forward cycle in the extended sense.

\bdfn[Local Termination Test, General Case]
For a cyclic $G$, 
we say that $G$ passes the \emph{Local $\pi$-Termination Test}, or LTT, if $G^\cv$
either has a balanced strict cycle, or both a forward 
cycle $\mathfrak f$ and a backward cycle $\mathfrak b$,
at least one of which is strict, and a  path from a node
of $\mathfrak f$ to a node of $\mathfrak b$.
\edfn

\begin{algo} (Closure Algorithm not presuming stability)
\end{algo}
\be
\item Build $\clos{\cal A}$.
\item
For each  cyclic $G$ in $\clos{\cal A}$, apply the Local $\pi$-Termination Test. \\
Pronounce failure (non-termination) if a graph that fails the test is found.
\item 
 If the previous step has completed, the MCS terminates.\smallskip
\ee

The proof of this algorithm much alike the proof of a similar
algorithm for the well-founded model~\cite{BA:mcs}, but rather tedious
and is omitted.

\subsection{Summary}
In this section we have presented a path-based condition for termination,
 its soundness and completeness, and sound and complete decision procedures 
based on it.
The condition comes in two flavours: there is a version for general MCSs and a version for stable ones. We concentrated on the latter.
Stabilization may be costly (the worst case incurs exponential blow-up), and hence the algorithms that do not require this
preprocessing may be more efficient in practice (at least, if all we require is to decide termination). We also included
Codish, Lagoon and Stuckey's algorithm, which is similar to the new algorithms proposed here.
The contribution of the new algorithms may be mostly
theoretical---a contribution to the understanding of MCS termination, as they are directly related to the path-based
criteria. In addition, we have
closed a gap in the completeness proof of~\cite{Codish-et-al:05}.  
In the sequel, an entirely different algorithm will be proposed, which constructs an explicit global ranking function.
If only the decision problem is of interest, the last (without stabilization) is probably the most efficient, except for simple cases
(such a simple case is SCT transition systems, since they are always stable).

\section{Full and Partial Elaboration}
\label{sec:elaborate}

In~\cite{BA:mcs}, some of the results were proved using the following observation: for a finite
number of variables, there are only finitely many orderings of their values. It is thus possible
to exhaustively list all possibilities and create an explicit representation of how transitions will
affect each one. This is called full elaboration of an MCS. 
It served two purposes in~\cite{BA:mcs}  (and so it will in the current paper): first, since it generates a stable
system, it gives an easy route to a decision procedure---one which is inefficient regarding time but can be used
to prove that the problem is decidable in polynomial space. The second purpose
was the algorithm to construct a global ranking function;
the fully-elaborated system has some structural properties that the construction was based on.
In this paper, too, we use full elaboration in that way.
The current section 
presents full elaboration as defined in~\cite{BA:mcs}, and infers PSPACE-completeness of the termination problem;
then, partial elaboration is introduced, as a tool to be used in the subsequent section.

\begin{ZvsWF}
The idea of full elaboration is taken from~\cite{BA:mcs} , and Theorem~\ref{thm:pspace}  is a straight-forward application, as it was
in the previous paper.
The idea and details of \emph{partial elaboration} are new definitions that are necessary for the new algorithm in the following section.
\end{ZvsWF}

\bdfn[full elaboration]
An MCS $\cal A$ is \emph{fully elaborated} if the following conditions hold:

\be
\item  Each state invariant
fully specifies the relations among all variables. That is, for
$i,j\le n$, one of the relations $x_i=x_j$, \ $x_i<x_j$ or \ $x_i>x_j$ is specified by $I_f$.

\item Each MC is closed under logical consequence.

\item Each MC in $\cal A$ is satisfiable.
\ee
\edfn

\newcommand{\lessoreq}{\left\{{<\atop =}\right\}}
Since the state invariant fully determines the relations among all
variables, we can \emph{re-index} the variables into sorted order,
so that the invariant becomes
\begin{equation}\label{eqn:order}
\textstyle x_{1} \lessoreq x_{2} \lessoreq\dots \lessoreq x_{n}.
\end{equation}
Of course, the re-indexing has to be incorporated also in
MCs incident to this flow-point, but this is straight-forward to do.
Indexing the variables in sorted order has some convenient consequences,
such as the having the property:

\bdfn \label{dfn:downwardclosure}
$G$ has the \emph{downward closure property}
if for all $k<j$, $G\vdash x_{i}{\ge} x_{j}'$ entails
 $G\vdash x_{i}{\ge} x_{k}'$.
\edfn

The number of possible orderings of $n$ variables, hence the maximum number of copies we may need
to make of any flow-point to achieve full elaboration, is the
\emph{the $n$th ordered Bell number} $B_n$, already mentioned.	
We denote the set of these orderings by $\bell_n$,
and assume that we fix some convenient representation so that 
``orderings'' can be algorithmically manipulated.

The algorithm of full elaboration follows almost immediately
from the definitions, and here it is, quoted from~\cite{BA:mcs}:
\begin{algo} (full elaboration) \label{alg-fe}
Given an MCS ${\cal B}$, this algorithm produces a fully-elaborated MCS $\cal A$, along with mappings
$\phi$ and $\psi$,  such that
$\mathcal A \sim_{\phi,\psi} \mathcal B$.
\end{algo}
\be
\item
For every $f\in F^{\cal B}$, generate flow-points 
$f_\pi$ where $\pi$ ranges over $\bell_n$.  
Define the variable renaming function $\psi_{f_\pi,f}$ so that
$\psi_{f_\pi,f}(i)$ is the $i$th variable in sorted order, according
to $\pi$. Thus,
$I_{f_\pi}$ will have exactly the form~(\ref{eqn:order}).
\item
Next, for every MC
$G:f\to g$ in $\cal B$, and every pair $f_\pi, g_\varpi$, create a size-change
graph $G_{\pi,\varpi}: f_\pi\to g_\varpi$ as follows:
\begin{enumerate}
\item
For every arc $x\to y\in G$, include the corresponding arc in $G_{\pi,\varpi}$,
according to the variable renaming used in the two $\cal A$ flow-points.
\item
Complete $G_{\pi,\varpi}$ by closure under consequences; 
unsatisfiable graphs (detected by the closure computation) are removed from the constructed system.\smallskip
\end{enumerate}
\ee

\noindent Since a fully elaborated system is stable, full elaboration allows termination to be checked by a closure
algorithm as previously described. However, full elaboration will often be costlier than stabilization by
fixed-point computation. Therefore, it is not worthwhile---unless we are interested in worst-case
complexity!  The complexity of the closure algorithm has an exponent of $n^2$, while full elaboration
only introduces an exponent of $n\log n$ and can subsequently be tested for termination by an algorithm of
the same exponent, as the next section will show. Full elaboration is also a convenient means to justify
the PSPACE upper bound.

\bthm \label{thm:pspace}
The MCS $\pi$-termination problem is PSPACE-complete.
\ethm

\bprf
It is known~\cite{leejonesbenamram01} that the SCT Termination problem
is PSPACE-hard, which also applies to MCS because SCT is a special case
(to reduce an SCT problem, which assumes well-foundedness, to a $\pi$-termination
problem, include a ``bottom" variable and constrain all others to be bigger.
This is like restricting the domain to the positive integers).

To show that the problem is in PSPACE, we will outline a non-deterministic
polynomial-space algorithm for the complement problem,
that is, non-termination.
The result will follow since (by Savitch's theorem)
$\mbox{coNPSPACE}=\mbox{NPSPACE}=\mbox{PSPACE}$.


Algorithm~\ref{alg-closure} can be seen
as a search for a counter-example---a cycle in the CFG that fails the test. 
The non-deterministic algorithm guesses such a cycle.
 In each step, it
adds a transition to the cycle while composing the transition's MC
with an MC that represents the CFG path traversed so far. Only this MC, along with the initial and current flow-point,
have to be maintained in memory.  Whenever the current flow-point is the same
as the initial one, the local termination test is applied.
If at some point, an unsatisfiable MC results, the algorithm has failed to find
a counter-example. Otherwise it continues until finding one.

Given an input MCS $\cal B$, we apply the Closure Algorithm to
the fully elaborated system $\cal A$ equivalent to $\cal B$.
To achieve the polynomial space bound, $\cal A$ is never constructed explicitly. 
In fact, to represent a flow-point $f_{\pi}$ of $\cal A$, we just maintain
the $\cal B$ flow-point $f$ along with the ordering $\pi$ (we do not re-index the variables). 
If the next flow-point is chosen to be $g_{\varpi}$ (both the flow-point and the ordering are chosen
non-deterministically), the MC $G_{\pi,\varpi}$ is computed on the spot.

It should be easy to see that the algorithm only needs access to the original MCS $\cal B$ and to an 
additional linear amount of memory.
\eprf

In the next section we will have a situation where we require information about the ordering
of variables which is not present in the given flow-point invariant, but we do not need the whole
ordering.  We gain efficiency by only performing a \emph{partial elaboration}, defined generally as follows:

\bdfn
Let $f$ be a flow-point in an MCS and let $I_1,\dots,I_k$ be conjunctions of order constraints that
are mutually exclusive and satisfy $I_1\lor I_2\lor \cdots I_k \equiv I_f$. 

A partial elaboration step applied to $f,I_1,\dots,I_f$ splits $f$ into $k$ flow-points, $f_i$ having the invariant
$I_i$, and accordingly replicates every MC from $f$ (or into $f$). Every such MC is consequence-closed under
the new source (or target) invariant, and eliminated if unsatisfiable.
\edfn

Again, it should be obvious that partial elaboration only refines the abstract program; it does not lose or
add possible runs, up to the renaming of flow-points. But it makes the program easier to analyse. 
As described, it makes the system possibly unstable. Stabilizing the system again may require additional
flow-point splitting.  For example, if we split $f$ according to the three possible relations among
$x_1$ and $x_2$, and there is an MC $f\to g$ with constraints $x=x' \land y=y'$, we will be forced to split
$g$ as well.  An important case in which we avoid this complication is SCT constraints, and in fact we can allow a bit more:

\bdfn
A \emph{semi-SCT} system is one in which there are no constraints of the form $x<y'$ or $x\le y'$. 
\edfn

This definition generalizes SCT because flow-point invariants are allowed.

\blem \label{lem:semi-SCT}
If $\cal A$ is semi-SCT and stable, applying a partial elaboration step to $\cal A$ results in a system
${\cal A}'$ which is stable as well.
\elem

\bprf
It is not hard to see that changing the flow-point invariant of $f$ in a semi-SCT system does not imply any
consequences for the ordering of variable values at other flow-points.
\eprf

In the next section we have to handle a situation where it is necessary to know
which variable has smallest value among $k$ unrelated variables, say $x_1,\dots,x_k$.
The reader may verify that it is possible to create $k$ mutually exclusive invariants such that in $I_i$,
$x_i$ is minimum (with some arbitrary breaking of ties).

\section{Ranking Functions for Integer MCSs}
\label{sec:grf}

In this section we develop the algorithm to construct global ranking functions for
Integer MCSs. The algorithm will process an MCS and either report that it is
non-terminating, or provide an explicit global ranking function. This problem has
been solved in the well-founded setting in~\cite{BA:mcs}, improving on a previous
solution for SCT~\cite{Lee:ranking}. Assuming stability, it is possible to solve the problem
in the integer domain by a reduction to the well-founded setting (creating difference variables).
This solution is not satisfactory, due to the potential squaring of the number of variables, figuring
in the exponent of the complexity. This motivates the development of a specialized algorithm.
The algorithm will achieve optimal results---in fact, a complexity similar to that obtained in the well-founded case.
Achieving this results required dealing with some complications that will be pointed out in the sequel.

In preparation, let us recall the definitions and result of~\cite{BA:mcs}.

\bdfn
A \emph{global ranking function} for a transition system $\cal T$
with state space $\mathit St$
is a function $\rho:{\mathit St}\to W$, where $W$ is
a well-founded set, such that $\rho(s) > \rho(s')$ for every $(s,s')\in {\cal T}$.

A ranking function for a MCS $\cal A$ is a 
ranking function for $T_{\cal A}$. Namely, it
satisfies $G\vdash \rho(s) > \rho(s')$ for every $G\in {\cal A}$.
\edfn

Remarks: (1) The qualifier \emph{global} may be omitted in the sequel, since we do not deal with
the notion of local ranking functions. (2) The definition highlights the role of the transition system 
$T_{\cal A}$. In fact, to find a ranking function for an MCS, we transform it into other 
MC systems that represent the transition system in a refined manner. Therefore the ranking function obtained
will be correct with respect to the original MCS.

\bdfn[vectors] \label{def-Vf}
Let $V$ be a set of variables. We define
$\vectors{V}$ to be the set of tuples 
$\vec v = \langle v_1,v_2,\dots\rangle$ of even length,
where every even position is a variable of $V$, such that every
variable appears at most once;
and every odd position is a non-negative integer constant.
\edfn

\bdfn \label{def-value}
The value of $\vec v \in \vectors{V}$ in program state $(f,\sigma)$,
denoted $\vec v \sigma$, is a tuple of integers obtained by
substituting the values of variables in $\vec v$ according to $\sigma$.
Tuples are compared lexicographically.
\edfn


\bthm[\cite{BA:mcs}] \label{thm:wfrf}
Suppose that MCS ${\cal B}$ is
terminating in the well-founded model, and has variables $V=\{x_1,\dots,x_n\}$.
There is a ranking function $\rho$ for $\cal B$ where
$\rho(f,\sigma)$ is described by a set of elements of $\vectors{V}$, each one associated with
certain inequalities on variables, which define the region where that vector determines the function value.
The complexity of constructing $\rho$ is 
$O(|{\cal B}|\cdot n^{2n+1})$.
\ethm

Here is an example, just to illustrate the form of the function:
\[ \rho(f,\sigma) = \left\{\begin{array}{cl}
        \langle 1,x_1,1,x_3\rangle\sigma & \mbox{if $x_1> x_2$} \\
        \langle 1,x_2,1,x_4\rangle\sigma & \mbox{if $x_1 \le x_2$}. 
\end{array}\right.
\]
Later, to simplify the presentation and the manipulation of such functions, we can omit $\sigma$
and write, for example
\[ \rho(f) = \left\{\begin{array}{cl}
        \langle 1,x_1,1,x_3\rangle & \mbox{if $x_1> x_2$} \\
        \langle 1,x_2,1,x_4\rangle & \mbox{if $x_1 \le x_2$}. 
\end{array}\right.
\]
We are thus dealing with functions that associate a symbolic tuple (or set of tuples selected by order constraints) to a flow-point.
Under the assumption of a well-founded domain, the set of tuples is also well-founded by the lexicographic order.

The theorem can be easily translated to a theorem for Integer MCS, based on using pairs of variables whose difference is
non-negative and forming a tuple over $\nats$ (so again we have well-foundedness). This is formalized next, and yields
our first (non-optimal, but simple to describe) solution.

\subsection{The Difference MCS}

 Let  $D = \{(i,j)\mid 1\le i < j \le n\}$. We introduce a variable $x_{(i,j)}$, with $(i,j)\in D$,
to represent each difference $x_j-x_i$.
 
\bdfn[difference MCS] \label{def:Delta}
Let $\cal A$ be a fully elaborated system, with $n$ variables  in each flow point,
indexed in ascending order of value.
The \emph{initial difference MCS} of $\cal A$ is an MCS ${\cal A}^\Delta_0$
where:

\be
 \item The set of flow-points is as in $\cal A$. The variables are  $V\cup W$, where
$V = \{x_1,\dots,x_n\}$ and $W = \{ x_{(i,j)} : (i,j)\in D\}$.
 
 \item For every flow point $f$, the state invariant $I_f^\Delta$ includes $I_f$ plus
any constraints that can be deduced from the constraints in $\cal A$, using the rule:
 $$x_i\le x_{\ell} \le x_{u} \le x_{j}  \Rightarrow x_{(\ell,u)} \le x_{(i,j)} .$$

 \item To every MC $G\in {\cal A}$, there is a corresponding MC, $G^\Delta$, in ${\cal A}^\Delta_0$.
It includes the constraints in $G$ plus
any constraints that can be deduced from the constraints of $G$ using the rules
\begin{align*}
x_i\le x'_{\ell} \le x'_{u} \le x_{j}  &\Rightarrow x'_{(\ell,u)} \le x_{(i,j)} \\
x_i <  x'_{\ell} \le x'_{u} \le x_{j}  &\Rightarrow x'_{(\ell,u)}  <  x_{(i,j)} \\
x_i\le x'_{\ell} \le x'_{u}  <  x_{j}  &\Rightarrow x'_{(\ell,u)}  <  x_{(i,j)} .
\end{align*}

 \ee
 
 \edfn

\noindent Observe that ${\cal A}^\Delta_0$ can be derived from ${\cal A}$ by a straight-forward,
polynomial-time algorithm. It is called \emph{initial} because in the algorithm presented later
it will be iteratively refined. But at this stage it suffices. We reproduce the observation from
Section~\ref{sec:conS}:
 
\begin{obs}
Satisfaction of Condition~\textrm{S} by $\cal A$ is equivalent to satisfaction of the SCT condition
by ${\cal A}^\Delta_0$, restricted  to the difference variables.
\end{obs}

This allows us to find a ranking function based on the difference variables. The original variables
($V$) will not be used in it. We keep them in the system, however, as an aid to the forthcoming, more efficient, algorithm.
Applying Theorem~\ref{thm:wfrf} yields

\begin{cor}
Suppose that MCS ${\cal B}$ is
$\pi$-terminating, and has variables $x_1,\dots,x_n$.
There is a ranking function $\rho$ for $\cal B$ where
$\rho(f)$ is given as a set of elements of $\vectors{W}$, each one associated with
certain inequalities on differences of variables. These inequalities define the region where that vector determines the function value.
There are at most $B_{n(n-1)/2}$ different vectors for any flow-point. 
The complexity of constructing $\rho$ is 
$O(|{\cal B}|\cdot \left({n(n-1)}/{2}\right)^{n(n-1)+1}) = O(|{\cal B}|\cdot n^{2n^2})$.
\end{cor}

To present the function in a readable way, we will replace the difference variables by expressions
$x_i-x_j$, so we obtain a function like this: (the example is only meant to illustrate the form)
\[ \rho(f) = \left\{\begin{array}{cl}
        \langle 1,x_2-x_4,1,x_3-x_4\rangle & \mbox{if $x_2-x_4 > x_2-x_3$} \\
        \langle 1,x_2-x_4,0,x_3-x_4\rangle & \mbox{if $x_2-x_4 \le x_2-x_3$} 
\end{array}\right.
\]
But this is an unsatisfactory result, because of the $n^2$ exponent and the fact that tuples may include up to $n(n-1)/2$
variable positions (or at least, this is the bound that the theorem gives).
The challenge tackled in the rest of this section is 
how to reduce the complexity to $n^{O(n)}$ and the number of variables in each tuple to $n-1$,
by solving the problem specifically rather than reducing it to the well-founded case.
This brings the upper bound close to lower bounds (for the length of the tuples, the worst-case
lower bound is matched exactly) based on the results of~\cite{BL:ranking2}.

\subsection{Overview}

The construction below follows the same basic outline as the one in~\cite{BA:mcs}, which is also similar
to other global ranking-function algorithms such as~\cite{BC:08:TACAS,ADFG:2010}.
The construction is iterative, based on the notions of a \emph{quasi-ranking function} and 
\emph{a residual transition system}.

\bdfn
Let $\cal T$ be a transition system with state space ${\it St}$.
A \emph{quasi-ranking function}
for $\cal T$ is a function $\rho:{\mathit St}\to W$, where $W$ is
a well-founded set, such that $\rho(s) \ge \rho(s')$ for every $(s,s')\in {\cal T}$.

The \emph{residual transition system} relative to $\rho$, denoted ${\cal T} / \rho$,
includes all (and only) the transitions of $\cal T$ which do \emph{not}
decrease $\rho$.
\edfn

The outline of the algorithm is: find a quasi-ranking function, generate a representation of the residual 
system, repeat as long as the system is not vacant (i.e., has transitions).
  If the quasi-ranking functions found are tuple-valued functions $\rho_1,\rho_2,\dots,\rho_k$,
it is easy to see that $\rho_1\cat \rho_2\cat \dots\cat \rho_k$ is a ranking function, where $\cat$ is tuple
concatenation, extended naturally to functions  (it is tacitly assumed that care is taken to maintain the
correct structure regarding even and odd positions in the tuples).

What are the quasi-ranking functions and how are they found? We have two major cases, depending on 
the connectivity of the control-flow graph. If it is not strongly connected, there are flow-points that
can only occur in a particular order, and we obtain the quasi-ranking function just by assigning suitable numbers to flow-points.
This is the easy case. The case of a strongly-connected system is the important one. In~\cite{BA:mcs}, it was shown that, thanks
to full elaboration, it is always possible to identify a single variable $x_{i_f}$ for each flow-point $f$ so that the function
$\rho(f) = x_{i_f}$ is quasi-ranking. The proof of the existence of such a variable, and the algorithm to find it, make essential use
of the fact that the $n$ variables are totally ordered by the flow-point invariant. In the
current setting, we have a total order on the original $n$ variables, but
 this does not induce a total order on the $n(n-1)/2$ \emph{difference variables}, and it is them
that will constitute the quasi-ranking functions. So, to obtain the results below, the main effort (given the results of~\cite{BA:mcs})
has been to find a way to deal with the partially-ordered set of difference variables without resorting to fully ordering them
(as that would create an $n^2$ exponent again).

\newcommand{\difMCS}{{\mathcal A}^\Delta}

Throughout the rest of this section, $\cal A$ denotes
a fully-elaborated MCS that satisfies Condition~S,
and $\difMCS_0$ the initial difference MCS.
The algorithm begins with $\difMCS_0$, and then, iteratively, refines and restricts it, while constructing
quasi-ranking functions, so that in general, the sub-algorithm for finding a q.r.f.~works on an MCS
that has already changed from the initial one.
We define \emph{a difference MCS} to be any MCS $\difMCS$ obtained from $\difMCS_0$ by a series of
partial elaboration steps, as defined in Section~\ref{sec:elaborate}, and restrictions.
Thus $\difMCS$ is one of a sequence of MC systems $\difMCS_j$, starting with $\difMCS_0$, so that
\begin{align*}
\difMCS_{2i+1} & \simeq_{\phi_i} \difMCS_{2i} \\ 
\difMCS_{2i+2} & \Subset \difMCS_{2i} 
\end{align*}
So, for all $j$, $\difMCS_j  \Subset_{\phi} \difMCS_0$ for an appropriate $\phi$.
The goal of each elaboration step is to allow for finding a q.r.f., and the subsequent restriction represents the residual transition
system relative to the current quasi-ranking function. Note that the elaboration steps do not involve a re-indexing of variables---this
is important since we rely on the indexing of the original elaborated MCS $\cal A$.

\subsection{Preparations for the construction}

We begin with a few definitions and properties that the algorithm relies on.
These definitions are taken (with some adaptation) from previous work on SCT, and therefore refer as \emph{threads} to what we have named
\emph{down-threads}. Since our analysis of the difference variables is SCT analysis---only down-threads
are considered---let it be understood that in the context of difference variables, a ``thread'' is a down-thread.

\bdfn[thread preserver]
Given difference MCS $\difMCS$, a mapping
$P:F^{\cal A} \to \powerset(D)$
is called a {\em thread preserver\/} of $\difMCS$
if for every $G^{\Delta}:f\to g$ in $\difMCS$ it holds that
whenever $\alpha\in P(f)$,
there is $\beta\in P(g)$ such that $G^{\Delta}\vdash x_{\alpha}{\ge}x_{\beta}'$.
\edfn

It is easy to see that the set of thread preservers
is closed under
union. Hence, there is a unique maximal thread preserver,
which we denote by $\mbox{MTP}(\difMCS)$.
Given a standard representation of $\difMCS$, 
$\mbox{MTP}(\difMCS)$ can be computed in linear time~\cite{BL:2006}.

We also need the following definition and results:

\bdfn[complete thread]
A thread in a given multipath is 
{\em complete} if it starts at the beginning of the multipath,
and is as long as the multipath.
\edfn

\blem \label{lem:finite}
Let $M_{\infty}$ be an infinite multipath. If every  finite prefix $M$ of $M_{\infty}$ contains a complete thread,
then $M_{\infty}$ contains an infinite thread.
\elem

\bprf
The proof is a straight-forward application of K\"onig's Lemma.
\eprf

\blem \label{lem:finiteM}
If a strongly connected MCS satisfies SCT, every finite multipath
includes a complete thread.
\elem

The proof is straight-forward and left out.

\bdfn[thread-safe]
We call a variable \emph{thread-safe} at flow-point $f$ if every finite multipath, starting at $f$,
includes a complete thread starting at that variable.
\edfn
\subsection{Finding a singleton thread-preserver} \label{sec:findpair}

The role of this part of the algorithm
is to single out a variable $x_{\alpha_f}$
for every flow point, such that these variables constitute a singleton
thread-preserver (i.e., $P(f)=\{\alpha_f\}$ be a thread preserver).
Observe that this induces a quasi-ranking function, namely $\rho(f) = \langle x_{\alpha_f} \rangle$.
We assume, until further notice, that we are dealing with a strongly-connected MCS; the complete algorithm
will provide for the non-strongly-connected case.

\let\pord=\subseteq
\let\pordrev=\supseteq

\bdfn[ordering of $D$]
We write $(i,j)\pord (i',j')$ for containment of the
interval $\{i,\dots,j\}$ in the interval $\{i',\dots,j'\}$. 
\edfn

Containment is a 
partial order, and moreover a semi-lattice with join operation $(i,j)\sqcup (i',j') =
(\min(i,j), \max(i',j'))$.

Recall that indexing the variables in ascending order of value, during
full elaboration, provided a useful \emph{downward closure} property. 
In particular, the lowest-numbered variable among a set of variables has the smallest value.
This was crucial in the construction of~\cite{BA:mcs}, and here, we give as a substitute
the following (more subtle) lemma for the difference variables.

\blem \label{lem:lattice}
Let $\difMCS$ be a difference MCS for a fully elaborated $\cal A$.
Each MC, $G^\Delta$, of $A^\Delta$ has the following properties: 
\begin{enumerate}[\em(1)]
\item \label{itm:lhs_closure}
 If $G^\Delta\vdash x_{\alpha}\ge x_{\beta}'$, then
$G^\Delta\vdash x_{\gamma}\ge x_{\beta}'$ for every $\gamma \pordrev \alpha$.
\item \label{itm:rhs_closure}
 If $G^\Delta\vdash x_{\alpha}\ge x_{\beta}'$, 
and $G^\Delta\vdash x_{\alpha}\ge x_{\gamma}'$,
 then
$G^\Delta\vdash x_{\alpha}{\ge} x_{\delta}'$ for every $\delta \pord
 \beta\sqcup\gamma$.
 \ee
\elem

\bprf We prove only the second claim (being a bit more involved).  Let $\alpha=(a_l, a_h)$,
$\beta=(b_l,b_h)$ and $\gamma=(c_l,c_h)$. The inequality $x_{\alpha}\ge x_{\beta}'$
means $x_{a_l}\le x_{b_l}'$ and $x_{a_h}\ge x_{b_h}'$ and similarly for the second inequality, so we have:
$$x_{a_l}\le x_{\min(b_l,c_l)} ; \quad x_{a_h}\ge x_{\max(b_h,c_h)}.$$
That is, $x_\alpha \ge x_{\beta\sqcup\gamma}$, which entails the conclusion.
\eprf

\blem \label{lem-tp-Delta}
Assume that $\difMCS$ is strongly connected and satisfies SCT in the difference variables.
For every $f$, let $S(f)$ be the set of indices of difference variables that are
thread-safe at $f$ in $\difMCS$. Then $S(f)$ is not empty for any $f\in F^{\difMCS}$
and  $S=\mbox{MTP}(\difMCS)$\footnote{This notation refers to a maximal thread preserver involving the difference
variables only.}.
\elem

\bprf
Let $M$ be any finite $\difMCS$-multipath starting at $f$.
Since $\difMCS$ satisfies SCT and is strongly connected,
there must be a complete thread in $M$, say starting at $x_\alpha$.
But then $x_{(1,n)}$ can also start a thread. It follows that $(1,n)\in S(f)$, so $S(f)$ is not empty.

We now aim to show that $S$ is a thread preserver.
Let $\alpha\in S(f)$,
and let $G^\Delta:f\to g$.
Let $H^\Delta$ be any MC with source point $g$, $M$ a finite multipath starting with $H^\Delta$ and  consider the
multipath $G^\Delta M 
 =G^\Delta H^\Delta\!\dots$. By definition of $S(f)$, it has
a complete thread that begins with an arc from $x_{\alpha}$, say
$x_{\alpha}\to x_{\beta_M}'$, followed by an arc of $H^\Delta$, say $x_{\beta_M}\to x_{\gamma_M}'$.
Let $B$ be the set of all such indices $\beta_M$ (where $M$ ranges over all multipaths starting with $H^\Delta$), and
$\kappa = \bigsqcup B$.
 Then $x_{\alpha}\to x_{\kappa}$ is an arc of $G^\Delta$
 by Lemma~\ref{lem:lattice}(\ref{itm:rhs_closure});
and by~\ref{lem:lattice}(\ref{itm:lhs_closure}) one can see that, for each $M$, $H^\Delta$ has an arc $x_{\kappa}\to x_{\gamma_M}'$.
Thus, $M$ has a complete thread beginning with
$x_{\kappa}$, and $\kappa$ is, by definition, in $S(g)$.
We have proved that $S$ has the thread-preservation property.
On the other hand, it is easy to see that if $P$ is a thread preserver such that $P(f)\neq\emptyset$, then
the variables indexed by $P(f)$ are thread-safe at $f$. Thus every thread preserver is contained in $S$, and we conclude that
$S=\mbox{MTP}(\difMCS)$.
\eprf

\blem \label{lem-tpmin-Delta}
Assume that $\difMCS$ is strongly connected and  satisfies SCT in the difference variables. 
Let $P$ be a thread preserver (in difference variables)
and suppose that 
for every $f\in F^{\difMCS}$, invariant $I_f^\Delta$ implies that a certain variable
$x_{\alpha_f}$ has smallest value among the variables of $P(f)$.
Then every
$G^\Delta:f\to g$ includes $x_{\alpha_f}\to x_{\alpha_g}'$. In other words,
$\{\alpha_f\}$ constitutes a singleton thread preserver.
\elem

\bprf
By the definition of a thread-preserver, each MC
$G^\Delta:f\to g$ must have an arc $x_{\alpha_f}\to x_{\beta}'$ for some ${\beta\in P(g)}$;
so by Lemma~\ref{lem:lattice}(\ref{itm:rhs_closure}), $G^\Delta$ includes $x_{\alpha_f}\to x_{\alpha_g}'$.
\eprf

Since the lemma applies to any thread preserver, the easiest implementation is to use the MTP, just because it
can be found efficiently.
The difficulty in applying the lemma is that the difference variables are only
partially ordered, so in a given set of variables there may be no variable that is necessarily of
smallest value.  In order to make the lemma applicable, we shall use partial elaboration.
Specifically, observe that given $P(f)$,
there can be at most $n-1$ candidates
for the variable of smallest value. This is so because in the poset of
intervals there is no antichain bigger than $n-1$.  Suppose that for flow-point $f$, the set of potential
smallest variables in $P(f)$,  $P_{\min}(f)$, has more than a single element. We will then
create a duplicate $f_{\alpha}$ of $f$ for each $\alpha\in P_{\min}(f)$, and add different constraints
to each one, 
such that in $f_{\alpha}$, $x_{\alpha}$ will be minimum, as required by the lemma.
This is a case of partial elaboration, as described in Section~\ref{sec:elaborate}, and due to the fact that the constraints on 
difference variables are semi-SCT, the duplication does not trigger
a cascade of duplications of other flow-points (Lemma~\ref{lem:semi-SCT}); we thus increase the size of the system at most
$(n-1)$-fold.  After this processing, Lemma~\ref{lem-tpmin-Delta} can be applied.

In the following lemma we formulate the conclusion in a somewhat generalized manner,
which will be useful later. Recall that $D$ is the set of indices of all difference variables.

\blem \label{lem:singletonTP}
Let $\difMCS$ be a strongly connected difference MCS, that satisfies SCT restricted to 
a certain subset of the difference variables at each flow-point $f$, given by $T(f)\subseteq D$,
such that $T(f)$ is closed under $\sqcup$.
Procedure $\procname{SingletonTP}({\difMCS},T)$ below finds a singleton thread preserver within $T$,
while (possibly) modifying the  MCS by partial elaboration.
The partial elaboration involves duplicating each point $f$ at most $n-1$ times.
\elem

\begin{algo} $\procname{SingletonTP}({\difMCS},T)$ \label{alg-singletonTP}
\end{algo}
\be 
\item
Compute $P = \mbox{MTP}(\difMCS,T)$, the MTP of $\difMCS$ restricted to the variables $T(f)$ for each $f$.
If empty, the procedure fails.
\item
For each $f\in F^{\difMCS}$,
identify the minimal elements of $P(f)$, based on the flow-point invariants. Suppose that 
there are $k>1$ minima, $x_{\alpha_1},\dots,x_{\alpha_k}$. Let $I$ be the invariant of $f$;
create mutually exclusive invariants $I_1,\dots,I_k$ such that $I_i$ includes $I$ with added constraints
$x_{\alpha_i} < x_{\alpha_j}$ for $j<i$, and $x_{\alpha_i} \le x_{\alpha_j}$ for $i\le i\le k$.
 Perform partial elaboration, replacing $f$ by $k$ copies with the respective invariants.
Choose $x_{\alpha_i}$ for the thread preserver at the $i$'th copy of $f$.
\ee
Note that the results of the procedure (unless it fails) are a possibly modified MCS and its thread preserver.


\subsection{Systems with frozen threads}

Having found a quasi-ranking function, our next step is
to modify to $\difMCS$ so that it expresses the residual system. This includes the following changes 
\be
\item Transitions in which the quasi-ranking function strictly decreases (that is, we have
$G^{\Delta}\vdash x_{\alpha_f} > x_{\alpha_g}$) are removed.
\item In
transitions where the quasi-ranking function was not known to decrease strictly (that is, we only had
$G^{\Delta}\vdash x_{\alpha_f} \ge x_{\alpha_g}$) it is now required not to change.\smallskip
\ee

\noindent We now recall another idea from~\cite{BA:mcs}, \emph{freezers}.
A freezer is a singleton thread-preserver where the values are ``frozen,'' that is, constrained not
to change.

\bdfn[freezer]
Let $\cal A$ be an MCS and
$C:F^{\cal A}\to \{1,\dots,n\}$ a function that
associates one original variable (technically, the index of one such variable) to to each flow-point.
Such $C$ is called a {\em freezer\/} 
if for every $G\in {\cal A}$, $G\vdash x_{C(f)} = x_{C(g)}'$.
\edfn

In our case, we have a thread of difference variables $x_{\alpha_f}$ that we wish to constrain not to 
change. Suppose that $G^\Delta \vdash x_{\alpha_f} \ge x_{\alpha_g}$, where $\alpha_f=(l,h)$ and $\alpha_g=(i,j)$.
In order that the differences before and after the transition be the same, we must have both
$x_l = x_i'$ and $x_h=x_j'$. Therefore, every arc like that, participating in the singleton thread-preserver
found, implies that two inequalities in \emph{original} variables are restricted to equalities, creating two
freezers.

Assuming that there still are cycles in the MCS, we need to
look for an additional quasi-ranking function.
We have to restrict the search in order not to find the same quasi-ranking function again (which, albeit
``frozen," is still a q.r.f.). This requires a bit further analysis.

The following observation comes easily from full elaboration:
\begin{obs}
Freezers are consistently ordered by the relations among their variables. That is, if $C_L$ and $C_H$ are freezers
and $I_f\vdash  x_{C_L(f)} < x_{C_H(f)}$ for some flow-point $f$, then this relation holds in every
flow-point and we write $C_L < C_H$. 
\end{obs}
Note that due to the re-indexing in
full elaboration,
the order relation among the freezers matches the relation among the \emph{indices} $C_L(f)$ and $C_H(f)$.
 The case of variables related by an equality constraint at a flow-point is less obvious but, in fact, if variable
 $x_1$, say, is ``frozen'' and $I_f \vdash x_1=x_2$, then $x_2$ is also ``frozen''. To make a long story short,
the algorithm below has the property that if two variables are constrained by equality, once one of them has been used
for a thread-preserver and thereafter put into a freezer, the other can safely (and advantageously) be ignored
(yet we ignore this situation in the description of the algorithm, to simplify presentation).


\blem \label{lem:pairfreeze}
Suppose that $\difMCS$ is strongly connected, and has some freezers,
among which $C_L$ is lowest and $C_H$ highest. And suppose that $\difMCS$ terminates.
Divide the indices of original variables, for every flow-point $f$, into three 
\emph{regions}: the lower region
$V_0(f) = \{1,\ldots,C_L(f)\}$,  the middle region $V_1(f) = \{C_L(f),\ldots,C_H(f)\}$,  and the upper region
$V_2(f) = \{C_H(f),\ldots,n\}$. 
Then every infinite $\difMCS$-multipath
has an approaching pair confined to one of the regions.
\elem

\bprf
Let $M$ be an infinite multipath of $\cal A$, $M=f_0\stackrel{G_1}{\to}f_1\stackrel{G_2}{\to}f_2\ldots$ 
Suppose that $M$ has an approaching pair: the low up-thread
$(x[k, l_k])_{k=0,1,\dots}$ and the high down-thread
$(x[k, h_k])_{k=0,1,\dots}$.

At least one of the low and high threads has to be infinitely often strict; 
let us suppose that it is the high thread (the other case is similar).
This thread cannot intersect any of the
the frozen threads infinitely often (or we would have an unsatisfiable section of the multipath,
contradicting Lemma~\ref{lem:stable=satisfiable}); so
in an infinite tail of the multipath (which is all that matters) it lies either
always above $C_H$, or below $C_L$, or between them; to avoid a trite case analysis,
let us pick just one
of the cases, and suppose that the thread uses variables above $C_H$, that is,
from $V_2$.
Now, we can let $C_H$ play the part of the low thread,
to obtain an approaching pair within the upper region ($V_2$).
\eprf



\begin{cor} \label{cor:thrd_in_region}
Under the assumptions of the last lemma, define regions of difference variables:
\begin{align}
D_0(f) &= \{(i,j)\mid i<j\le C_L(f)\};   \label{eq:D0}\\
D_1(f) &= \{(i,j)\mid C_L(f)\le i<j\le C_H(f)\}; \\
D_2(f) &= \{(i,j)\mid C_H(f)\le i<j\};  \label{eq:D2}
\end{align}
then in every infinite $\difMCS$-multipath there is an infinitely-descending thread of difference variables
within one of these regions.
\end{cor}

\blem \label{lem:pair_in_region}
Under the assumptions of the last lemma,
there is a region $D_r$  that contains an infinite thread (corresponding to a weakly approaching pair
in $V_r$) in every infinite $\difMCS$-multipath.
\elem

Note the change in the order of quantification: the region is selected before the multipath,
 at the expense of not guaranteeing strict descent in every multipath.

\bprf 
Assume to the contrary that for each of $r=0,1,2$ there is an infinite multipath in which Region $D_r$
contains no infinite thread.  By Lemma~\ref{lem:finite}, there are finite multipaths
$M_0$, $M_1$, $M_2$ such that $M_i$ has no complete thread in
region $i$.  Since our MCS is strongly connected, one can form a multipath
$(M_0\dots M_1\dots M_2\dots)^\omega$, which will have no infinite thread
in any of the regions, contradicting Corollary~\ref{cor:thrd_in_region}.
 \eprf

Let $r$ be such that there is always an infinite thread in $D_r$.
We can use the strategy of the previous subsection to find a singleton thread preserver,
thus making progress in our construction.  But is it progress? There is a pitfall:
in the case of the middle region, our procedure may find a pair of variables that was already frozen 
(they constitute a weakly approaching pair, but they really never approach).
However, the middle region can be treated in a special way that avoids the pitfall and is also
more efficient.

\blem \label{lem:middle_region}
Under the assumptions of Lemma~\ref{lem:pairfreeze},
An infinite multipath $M$ contains an approaching pair within the
middle region ($V_1$) if and only if
this region contains a thread---either a down-thread or an up-thread---which is infinitely often strict. Such a thread is disjoint from the two freezers delimiting this region.
\elem

\bprf The non-trivial implication is the ``if," but it is also quite easy: suppose that the region
contains an infinitely-often strict down-thread (up-thread). It can be complemented to an approaching pair
by using $C_L$ ($C_H$) for the other thread. 
\eprf

From this lemma we conclude that the middle region can be reduced to the following two subsets of difference variables:
\begin{align}
D_1^L(f)  &=  \{ (C_L(f),j) \mid C_L(f) < j < C_H(f) \text{ such that $x_j$ is not frozen in $f$} \}  \label{eq:D1L}\\
D_1^H(f) &= \{ (i,C_H(f)) \mid C_L(f) < i < C_H(f) \text{ such that $x_i$ is not frozen in $f$} \}.  \label{eq:D1H}
\end{align}
Both sets of intervals are closed
under $\sqcup$ and Lemma~\ref{lem:singletonTP} applies.
Moreover,  these sets of intervals are totally ordered,
which means that no elaboration steps will be needed when looking for a singleton thread-preserver
among them.

\subsection{Putting it all together}


\begin{algo} (ranking function construction for ${\difMCS}$) \label{alg-genA}
\end{algo}
Assumes that $\difMCS$ is a difference MCS.
The system may also be adorned
with any number of freezers.
If $\difMCS$ terminates,
a ranking function will be returned. Otherwise, the algorithm will fail.
\be 
\item
List the SCCs of ${\difMCS}$ in reverse-topological order. For each $f\in F^{\difMCS}$,
let $\kappa_f$ be the
position of the SCC of $f$.
Define $\rho(f) = \langle\kappa_f\rangle$.

If all SCCs are vacant (contain no transitions), return $\rho$.

\item
For each SCC $C$, compute a ranking function $\rho_C$ by applying the next algorithm to the
component (separately).
Let $\rho' = \bigcup \rho_C$. Return $\rho\cat \rho'$. 
\ee

\begin{algo} (for a strongly connected ${\difMCS}$) \label{alg-sccA}
\end{algo}
\be 
\item
If no freezers are associated with $\difMCS$,
run $\procname{SingletonTP}({\difMCS}, D)$ and proceed to Step~3.
\item
If some freezers are associated with $\difMCS$,
let $C_L$ and $C_H$ be the lowest and highest freezers\footnote{A case in which there is only one freezer
is acceptable. $C_L$ and $C_H$ are then the same thread and Step (2a) is skipped.}.
For every flow-point
$f$, create the sets of difference variables $D_0(f)$, $D_1^L(f)$,  $D_1^H(f)$, $D_2(f)$,
as defined in Equations~(\ref{eq:D0}--\ref{eq:D1H}).
\be
  \item
  Run $\procname{SingletonTP}({\difMCS},D_1^x)$ for $x=L$ and then $H$.
If one of these calls succeeds,
proceed to Step~\ref{stp:freeze}.
  \item
  Run $\procname{SingletonTP}({\difMCS}, D_0)$. If successful,
proceed to Step~\ref{stp:freeze}.
 \item
  Run $\procname{SingletonTP}({\difMCS}, D_2)$. If successful, proceed to Step~\ref{stp:freeze}. 
If not, the algorithm fails.
\ee
\item \label{stp:freeze}  (Create residual system) Let $P$ be the thread preserver found.
     For every graph $G^\Delta:f\to g$, if it
     includes $x_{P(f)} > x_{P(g)}'$,
     delete the graph from ${\difMCS}$.
     Otherwise, retain the graph. Also include
\footnote{We use the projections $\pi_1,\pi_2$ to map indices from $D$ to their components.}
 $C_1(f) = \pi_1 P(f)$, $C_2(f)= \pi_2 P(f)$ as freezers associated with $\difMCS$.
\item
 For every $f$,  let $\rho(f) = \langle x_{P(f)}\rangle$.
\item
If ${\difMCS}$ is now vacant, return $\rho$.
Otherwise,
compute a ranking function $\rho'$ recursively for (what remains of) ${\difMCS}$,
using Algorithm~\ref{alg-genA}, and return $\rho\cat \rho'$. 
\ee

Recall that $\tilde O(f)$ is a shorthand for $O(f\cdot \log^{O(1)} f)$. Thus $\tilde O(n^n)$ is asymptotically dominated by
$n^n$ times a polynomial in $n$.
 For an MCS $\cal A$, let $|{\cal A}|$ denote the number of abstract
 transitions (MCs) in ${\cal A}$ (without loss of generality, $|{\cal A}|
\ge |F^{\cal A}|$).

\bthm
Let ${\cal A}$ be a fully-elaborated,
$\pi$-terminating MCS, with $n$ variables per point.
Algorithm~\ref{alg-genA}, applied to $\difMCS$, produces
a ranking function $\rho: F^{\cal A}\to \vectors{W}$
where each vector includes at most $n-1$ difference variables.
The complexity of construction of $\rho$ is
$\tilde O(|{\cal A}|\cdot n!)$.
\ethm

\bprf
Assuming that the correctness of the algorithm has been justified convincingly enough, we now discuss
the complexity. One should consider the effect of
partial elaborations by $\procname{SingletonTP}$. Such a step may
make up to $n-1$ copies of every flow-point. However, each time it is performed,
at least one new freezer is subsequently created (in the first time, two freezers).
Therefore, this multiplication of flow-points can occur at most
$n-1$ times, leading to the upper bound on the length of the vectors.
Since the number of variables participating in the search for quasi-ranking functions diminishes in each iteration,
we obtain a bound of $|F^{\cal A}|\cdot n!$  on the size of the resulting expression,
which associates a vector with each of the flow points of the mostly elaborated system obtained.
The running time
is further multiplied by a (low order) polynomial expressing the complexity of the procedures
at each level of the recursion.
\eprf

\bthm
Let ${\cal B}$ be a
$\pi$-terminating MCS, with $n$ variables per point.
A ranking function $\rho$ for $\cal B$ where
$\rho(f)$ is given by a case expression with inequalities among differences for
guards; the value in each case is given by a vector in $\vectors{W}$,
that includes at most $n-1$ difference variables.
The complexity of construction of $\rho$, as well as the size of the expression, are
$\tilde O(|{\cal B}|\cdot 2^{n}(n!)^2)$.
\ethm

\bprf
First, fully elaborate $\cal B$, yielding an MCS of size at most $|{\cal B}|\cdot B_n \le |{\cal B}|\cdot 2^n(n!)$, then use the last theorem.
\eprf

The algorithm could, in principle, be used just to determine if a system is terminating, and with an
exponent of $O(n\log n)$, better than the $O(n^2)$ exponent given in Section~\ref{sec:walks}. However, the use
of full elaboration makes it unattractive in practice  because it blindly generates the worst case (all possible ordering of variables)
for every input instance. In contrast, the algorithms of Section~\ref{sec:walks} will often perform much better than their
worst-case behaviour.

\section{Rooted Versus Uniform Termination}
\label{sec:rooted}

Up to this point, the notion of termination used was 
{\em uniform termination}, which means that there must be no cycles in the 
whole state space of the modeled transition system. Practically, what we usually
require is \emph{rooted termination}, when only computation paths beginning at a given
initial point $f_0$ are considered.

\begin{ZvsWF}
There is no difference between the treatment of this subject in the well-founded case and here, but the examples below
should illustrate that the issue is doubly important in the current setting.
\end{ZvsWF}

\noindent
Here is a little C example to show the importance of rooted termination:

\begin{program}
if (x<0) {
    while (y > 0)  y = y+x 
}	
\end{program}

\noindent Consider an abstraction that represents the command 
\pgt{y=y+x} by three parallel MCs, as discussed in Example~\ref{ex:addition}.
It can be shown to terminate when only paths from the
top of the program are considered.
The example could also be solved by a preprocessing that calculates
state invariants, such as $\pgt{x}<0$ inside the \pgt{while}, as is often done in program analysers.
But this can get complicated: the following example
would require an invariant that specifies
the dependence of variable \pgt{b} on \pgt{x}.
This kind of invariant 
that is not found in common invariant generators (namely those that describe a state by a conjunction
of linear constraints).
However with rooted termination, no  invariants are
necessary other than the direct translation of the conditionals to monotonicity constraints.

\begin{program}
if (x<0)  b=1
    else  b=0;
if (x<>0)
    while (y > 0)  {
        if (b) y = y+x; 
          else y = y-x;
}	
\end{program}

\noindent We conclude that it is desirable, practically, to account for rooted termination.
Up to this point, this was avoided only in favor of simplicity of presentation. However,
it is very easy to do: with a stable system, unreachable states will be
represented by flow-points that are inaccessible from $f_0$, due to Lemma~\ref{lem:stable=satisfiable}.

Thus, all that is necessary is to remove inaccessible parts of the CFG, or better yet, never generate
them in the first place, by creating the stabilized (or fully elaborated) system in the manner of a
graph exploration (say, DFS) starting at the initial point, only covering the reachable state space. This 
can occasionally have a significant effect on efficiency (in particular with full elaboration), as
confirmed by our experience with implementing full elaboration~\cite{ArielSnir}.

\bex \label{ex:rootedDec}
To conclude this section, here is another example which illustrates the effectiveness of the MCS abstraction in expressing disjunctions,
besides the need for rooted termination.
Disjunction is used in expressing the condition \pgt{x != 0}, as well as representing the command \pgt{x := x-1}.

\medskip
\begin{program}
assert{ x > 0 }
while (x != 0)  x := x-1 
\end{program}
\medskip

\noindent Abstraction: unlike previous examples, we will not merge this time the loop guard with the loop body, in order to clarify
the transformation. We will thus have three flow-points, $0$ (initial),  $1$ (loop header) and $2$ (loop body).
\begin{align*}
G_1:0\to 1&:\quad
 \pgt{x}>\pgt{0}\land \pgt{x}=\pgt{x'}\land \pgt{0}=\pgt{0}'\\
G_2:1\to 2&:\quad
 \pgt{x}>\pgt{0}\land \pgt{x}=\pgt{x'}\land \pgt{0}=\pgt{0}'\\
G_3:1\to 2&:\quad
 \pgt{x}<\pgt{0}\land \pgt{x}=\pgt{x'}\land \pgt{0}=\pgt{0}'\\
G_4:2\to 1&:\quad
 \pgt{x}>\pgt{0}\land \pgt{x}>\pgt{x'}\land \pgt{x'}\ge\pgt{0'}\land \pgt{0}=\pgt{0}'\\
G_5:2\to 1&:\quad
 \pgt{x}\le\pgt{0}\land \pgt{x}>\pgt{x'}\land \pgt{x'} <\pgt{0'}\land \pgt{0}=\pgt{0}'
\qquad \qedhere\\
\end{align*}
\eex

\section{Some  Related work}
\label{sec:rw}

The field of termination analysis is well developed and it is infeasible to survey it extensively here. This section will point out
some works that are related inasmuch as they base a termination analysis on the behavior of integer variables.
The following questions are asked when considering such works: 

(1)  Does the proposed algorithm or tool work on an abstract
transition system, or on concrete programs? Clearly the case that can be best related to the current work is the former. Therefore, much work of the second kind is ignored here; but it is not hard to find.

(2) What abstraction is used? In particular, are monotonicity constraints used?  

(3) What is the main technique?  Is it complete for the given abstraction? (It is also possible to ask if there 
would have been completeness,  had the method been applied to MCS.  I have tried to answer this question,
though some of the methods are not fully described in the publications, so it is difficult to be precise.)

We start with those works which mostly resemble the current paper.

\begin{enumerate}[(1)] \itemsep 1ex
\item
Manolios and Vroon \cite{MV-cav06} describe a termination analysis 
implemented in the ACL2 theorem prover. It works on concrete programs using the SCT abstraction.
It handles termination arguments involving integers by introducing
the difference of two concrete integer variables as an abstract variable, when it can be determined
(using theorem proving techniques applied to the source program)
to be non-negative. Of course, it may also include a single integer variable if it is determined to be lower-bounded.
Since concrete programs are the subject, no completeness claim is made. It would be possible to represent an
MCS as a program and apply the tool, but completeness is still unlikely because,
as previously remarked, this reduction to SCT 
has to be combined with stabilization to achieve completeness.

\item
Avery~\cite{Avery:06} describes a tool to analyze C programs by first abstracting them to a constraint transition system
of the following form: transition constraints use inequalities ($\ge$, $>$, $\le$, $<$) to relate source and target variables;
flow-point invariants are polyhedral ones, that is, conjunctions of linear inequalities in state variables and constants.
This is clearly a generalization of MC transition systems, however one expressive enough to represent counter programs,
which means that termination is undecidable. The (sound but incomplete) algorithm is based on closure computation, 
where the composition operation used to form the closure takes into account only the variants (source-to-target relations, which are
MCs) and not the invariants.
For each idempotent graph in the closure, the invariants are taken into account when deciding which variables would imply
termination if they descend. 

Clearly, the algorithm could be applied to MCs. What bars it from achieving completeness in this case is the fact that the control-flow
graph is not refined (i.e., no stabilization).  Here is an example to illustrate this limitation (the program consists of a single loop
with the following description)
$$ \pgt{y}>\pgt{w} \land 
 \pgt{x}>\pgt{x}'\land \pgt{x}>\pgt{y}'\land \pgt{w}\le \pgt{z}'\land \pgt{w}\le\pgt{w}' 
 $$
Note that the only invariant which is valid whenever the transition is entered is $ \pgt{y}>\pgt{w}$, which is not very helpful.

\item
\textsc{Termilog}~\cite{LS:97} was a termination analyzer for Prolog, that made use of an abstraction to 
monotonicity constraints and  a closure computation plus a local test, which is sound but
incomplete for the MC constraint domain, as pointed out in~\cite{Codish-et-al:05}. 
The variables of the abstraction represent certain \emph{norms} of symbolic terms in the Prolog program.
They are, therefore, non-negative integers and the termination proof only looks for descent
towards zero. 
 We should remark, however that they do not use ``abstract compilation" as in our examples.
 Instead, an abstract interpreter is used to compute the closure set. This may be 
more precise with respect to the semantics of the subject program, see~\cite{HJP:2010}.

In~\cite{DLSS:2001}, Dershowitz et al.\ reformulated the principles underlying the usage of monotonicity constraints in
\textsc{Termilog}, and also proposed an extension to handle programs with integer variables
and arithmetics.
Their proposal is based on creating a ``custom-tailored" domain for abstraction of the integer variables,
based on constraints extracted from the program.
Their algorithms are Prolog-specific, and do not analyze a contraint-based abstract program,
but it seems that it could be applied to MC transition systems (appropriately
represented) and that the proposed abstraction process may actually compute a (partial) elaboration and obtain a stable system. 
It will further attempt to prove termination for
every cyclic MC by ``guessing'' a local ranking function. The functions they propose to ``guess" are differences ($x_i-x_j$), which
we know to suffice
for Idempotent cyclic MCs (See Definition~\ref{def:LTT1}: the function is $x_h-x_l$).
However, such functions do not suffice for all cyclic MCs \cite{Codish-et-al:05}.  We conlcude that it would be desirable to
apply the idempotence-based algorithm with their framework (which, in fact, they do in the part that deals with symbolic variables and norms).

\item
\textsc{Terminweb}~\cite{CT:99} is a termination analyzer for Prolog. It uses a procedure which tries to prove
termination of an abstract program, and there are two kinds of abstractions used. In the first, transitions
are described by polyhedra. In the second, by monotonicity constraints. In both cases, 
the data are non-negative integers and the termination proof sought is based on descent towards zero.
As for the methods, closure computation and
a local termination test are used; in the polyhedral case, the closure is approximated (using widening) for otherwise
it might be infinite. For the monotonicity-constraint abstraction, the closure computation is precise, but the
local termination test is incomplete, as pointed out in~\cite{Codish-et-al:05}.

\item
Mesnard and Serebrenik~\cite{MS:08} show that for abstract programs with transitions defined by polyhedra (conjunctions of 
linear inequalities), when the data are rationals or real numbers, it is possible to determine in polynomial time (using linear programming)
whether there is a global ranking function  that associates an affine combination of the variables with each flow-point.
This is an extension of the idea previously presented by Sohn and van-Gelder~\cite{SohnVanGelder:91}.
The existence of such a function is, of course, a sound (but incomplete) criterion for termination, and restricting the data to integers
maintains its soundness.

\item
The \textsc{BinTerm} analyzer of Spoto, Mesnard and Payet analyzes abstract programs and serves
as a back-end to the Java Bytecode termination analyzer \textsc{Julia}~\cite{SpotoMP09}.
The abstraction that it uses is a transition system with polyhedral constaints, 
and it applies a selection of strategies, which it tries one by one. The first two correspond to the two methods of
\textsc{Terminweb}, slightly modified since the domain is now the integers.
Thus, both for polyhedral transitions and for MCs, its local test is based on a search for an affine ranking function.
In the MC domain, such a test would be complete in the stable case, but is not complete in general.
The third method used is the method of Mesnard and Serebrenik.


\item
Col{\'o}n and Sipma~\cite{CS:02} is representative of a series of works that ostensibly target imperative programs,
but work, in fact, on a constraint-transition system with linear (affine) constraints (the domain may be assumed to be
the rationals, the reals or the integers) and search for global ranking functions of the lexicographic-linear type, using
linear programming techniques.

 Alias et al.~\cite{ADFG:2010} use the same general approach, but their class of ranking
functions is more general (specifically, each q.r.f.~associates a linear expression with every flow-point, whereas in~\cite{CS:02},
a q.r.f.~is a single expression throughout a SCC).  The lexicographic approach is, of course, more general than just looking for
a single affine global ranking function (as in~\cite{MS:08}), but still does not guarantee completeness for MCSs, where
the ranking function sometimes has to depend on the order relations of the variables, and so is not linear
 (a simple example is a ranking function $\min(x,y)$). 

Both methods rely on polyhedral invariants associated with
 a flow-point (in~\cite{ADFG:2010}, they are part of the abstract program, and would be generated by a front-end; in~\cite{CS:02},
 they are recomputed in each stage of the algorithm, which may improve its precision).

\item
Noting that monotonicity constraints are a special case of polyhedral constraints, it is natural to look for other
interesting subclasses, possibly richer than  monotonicity constraints.
 The class of \emph{difference constraints}  is defined by
constraints of the form $x-y\le c$. Termination of such constraint transition systems is shown
undecidable in~\cite{BA:delta}. Decidability in PSPACE is proved for a restricted subclass, called
\emph{fan-in free $\delta$SCT}.  This class is incomparable to MCSs (their intersection is fan-in free SCT).

\item
There are several published works that address a special subclass of constraint transition systems: simple loops, namely
transition systems with only one control location (flow point). Moreover, some of them consider a single-path loop,
consisting of one abstract transition only. This appears like a far-fetched restriction, but nonetheless, such simple
loops can be complex enough to merit theoretical interest, and practically, an algorithm to decide termination of such 
a loop can be used as the local test in a closure-based algorithm, or an algorithm based on counter-example based search (see~\cite{CPR06} for a well-known example of the approach).

Single-path loops, represented with polyhedral constraints, are handled using linear ranking functions in~\cite{PR:04}.
Bradley, Manna and Sipma~\cite{BMS:LexLinear}  extend this to a multi-path loop,
and show how to find a lexicographic-linear ranking function. 	In~\cite{BMS:Polyranking},
the approach is generalized so that the components of the ranking tuple are not required to be
quasi-ranking functions, but only ``eventually'' quasi-ranking, that is, they may ascend initially but must
eventually descend (as in Example \ref{ex:addition}).\smallskip
\end{enumerate}

\noindent A few of the works mentioned also generate global ranking functions. \cite{ADFG:2010} generates 
lexicographic-linear ranking functions. This is also the case with~\cite{CS:02}, although they are not
explicit about it, and their class of functions is more restricted. In~\cite{MS:08}, the class is restricted
to affine functions. All the works mentioned for analyzing simple loops are ranking-function based,
but only those in~\cite{BMS:LexLinear,BMS:Polyranking} may be truly called ``global" since they apply
to all the paths of a multi-path loop.

\section{Conclusion and Research Questions}
\label{sec:conclusion}

We studied the MCS abstraction, an appealing extension of the Size-Change
Termination framework, that can be used for termination analysis in the integer domain.
We showed how several elements of the theory developed in the well-founded model can
be achieved in the integer case: sound and complete termination criteria,
closure-based algorithms and 
the construction of ranking functions in singly-exponential time.
Global ranking functions may be useful for certified
termination~\cite{krauss07,CCFPU:07,KoprowskiZantema08}
and cost analysis~\cite{AAGP:sas08,ADFG:2010}, and the complexity achieved here is better
than what has been published before for SCT.

Hopefully, this paper will trigger further research, moving
towards the practical application of the theory presented. Some of the systems mentioned in the last section can
gain an increase in precision by incorporating a complete decision procedure for monotonicity constraints,
and it is encouraging that abstraction of a concrete program to monotonicity constraints either exists already in these
systems or can be added with very little effort, typically because \emph{richer} domains are already used (such as affine relations).

The algorithms in this article were aimed at getting the theoretical results with
a minimum of complications. They can certainly be improved in practice (as discussed in the
conclusion of~\cite{BA:mcs}).

Cases like Example~\ref{ex:int_and_bool} suggest that it may
be worthwhile to treat Boolean variables as such, so that they do not 
get entangled with the integer variables in the course of elaboration,
creating an unnecessary combinatorial explosion. A better idea 
is to extend the MCS abstraction to include Boolean variables
and extend the termination criteria and algorithms to account for them
precisely.  This may be a useful extension in practice, and moreover, 
it allows for adding information that does not come straight-forwardly from
the program, in the form of ``invented'' Boolean variables---leveraging
abstraction techniques used in the area of model checking.

Here are a few other directions for extension of this work:
\be
\item
Investigating extensions of the constraint domain, particular to the integers (i.e., not appropriate for
general SCT).
An example is difference constraints (mentioned in the last section).

\item
Proceeding from termination analysis to analysis of a program's complexity \cite{AAGP:sas08,ADFG:2010}.

\item
The idea of using multiple abstractions (one may
speak of abstractions of varying refinement) in a single tool in quite enticing.
One can also consider an abstraction-refinement
loop~\cite{CPR06} which allows for eliminating spurious counterexamples by specialized tools, while
using size-change analysis as a backbone.
\ee

\subsubsection*{Acknowledgments.} The author thanks
the APL group at DIKU  (the Computer Science department at the University of Copenhagen),
where part of this work was done, for hospitality,
and the anonymous referees, whose 
thorough reviews and suggestions contributed 
significant improvements to this paper.

\bibliographystyle{plain}
\bibliography{sct}

\end{document}